\documentclass[entropy,article,accept,pdftex,moreauthors]{Definitions/mdpi}

\firstpage{1} 
\pubvolume{1}
\issuenum{1}
\articlenumber{0}
\pubyear{2025}
\copyrightyear{2025}
\externaleditor{Reik Donner}
\datereceived{18 October 2024} 
\daterevised{26 February 2025} 
\dateaccepted{19 March 2025} 
\datepublished{ } 
\hreflink{https://doi.org/} 
\usepackage{xcolor} 

\Title{\highlighting{A Revisit} 
 of Large-Scale Patterns in Middle Stratospheric Circulation Variations}

\TitleCitation{A Revisit of Large-Scale Patterns in Middle Stratospheric Circulation Variations}

\Author{\hl{Ningning Tao} 
 $^{1}$, Xiaosong Chen $^{1,2}$*, Fei Xie $^{1}$* , Yongwen Zhang $^{3}$ , Yan Xia $^{1}$, Xuan Ma $^{1}$, Han Huang $^{1}$\linebreak and Hongyu Wang $^{1}$}

\AuthorNames{Ningning Tao, Xiaosong Chen, Fei Xie, Yongwen Zhang, Yan Xia, Xuan Ma, Han Huang and Hongyu Wang}

\AuthorCitation{\hl{Tao, N.; } 
Chen, X.; Xie, F.; Zhang, Y.; Xia, Y.; Ma, X.; Huang, H.; Wang, H.}

\address{%
$^{1}$ \quad School of Systems Science, Beijing Normal University, Beijing 100875, China; \linebreak  \hl{ningningtao@mail.bnu.edu.cn (N.T.); xiayan@bnu.edu.cn (Y.X.); maxuan@mail.bnu.edu.cn (X.M.); 202431250013@mail.bnu.edu.cn (H.H.); hongyuwang@mail.bnu.edu.cn (H.W.)} 
\\
$^{2}$ \quad Institute for Advanced Study in Physics \hl{and} 
 School of Physics, Zhejiang University, Hangzhou 310058, China\\
$^{3}$ \quad Data Science Research Center, Faculty of Science, Kunming University of Science and Technology, \linebreak Kuming 650500, China; \hl{zhangyw@kust.edu.cn}
}
\corres{Correspondence: chenxs@bnu.edu.cn (X.C.); xiefei@bnu.edu.cn (F.X.)}

\abstract{Variations in stratospheric atmospheric circulation significantly impact tropospheric weather and climate. Understanding these variations not only aids in better prediction of tropospheric weather and climate but also provides guidance for the development and flight trajectories of stratospheric aircraft.
Our understanding of the stratosphere has made remarkable progress over the past 100 years. However, we still lack a comprehensive perspective on large-scale patterns in stratospheric circulation, as the stratosphere is a typical complex system. To address this gap, we employed the eigen microstate approach (EMA) to revisit the characteristics of zonal wind from 70\highlighting{--}
10 hPa from 1980 to 2022, based on ERA5 reanalysis data. Our analysis focused on the three leading modes, corresponding to variations in the strength of the quasi-biennial oscillation (QBO) and the stratospheric atmospheric circulations in the Arctic and Antarctic, respectively. After filtering out high-frequency components from the temporal evolutions of these modes, a significant 11-year cycle was observed in the Antarctic stratospheric atmospheric circulation mode, potentially linked to the 11-year solar cycle. In contrast, the Arctic stratospheric atmospheric circulation mode showed a 5\highlighting{--}6-year cycle without evidence of an 11-year periodicity. This difference is likely due to the timing of polar vortex breakdowns: the Antarctic polar vortex breaks up later, experiencing its greatest variability in late spring and early summer, making it more susceptible to solar radiation effects, unlike the Arctic polar vortex, which peaks in winter and early spring. The fourth mode exhibits characteristics of a Southern Hemisphere dipole and shows a significant correlation with the Antarctic stratospheric atmospheric circulation mode, leading it by about two months. We designed a linear prediction model that successfully demonstrated its predictive capability for the Antarctic polar vortex.}


\begin{document}


\section{Introduction}

The stratosphere, located approximately 10 to 50 km above the Earth's surface and containing around 17\% of the Earth's atmospheric mass, is crucial to human life.  As early as 2001, Baldwin pointed out that anomalies in stratospheric circulation can descend to influence the troposphere and serve as predictive signals for weather and climate \citep{baldwin2001stratospheric}. Today, stratosphere processes---along with their coupling with the troposphere---play a significant role in sub-seasonal to seasonal weather predictions \citep{domeisen2020role}. 

Our understanding of the stratosphere has evolved significantly over the past 100 years \citep{baldwin2019100}. In the early 20th century, the existence of the stratosphere and the ozone layer was first discovered. With advancements in observational technology, an increasing number of stratospheric processes have been identified. Concepts and theories explaining these processes were subsequently developed, including the Brewer--Dobson circulation \citep{brewer1949evidence,dobson1956origin,butchart2014brewer}, the sudden stratospheric warmings (SSWs) \citep{scherhag1952explosionsartige,matsuno1971dynamical,labitzke1981stratospheric,baldwin2021sudden}, the quasi-biennial oscillation (QBO) \citep{ebdon1960notes,reed1961evidence,lindzen1968theory,holton1972updated,baldwin2001quasi}, the wave--mean flow interaction \citep{eliassen1961transfer,andrews1978exact,boyd1976noninteraction}, the parameterization of waves in climate models \citep{marks1995three,alexander1997model,alexander1999spectral,geller2013comparison}, the stratosphere--troposphere coupling \citep{domeisen2020role,baldwin2024tropospheric}, the concept of ``mean age of air'' \citep{hall1994age}, the chemical reactions of stratospheric ozone \citep{dobson1931photoelectric,bates1950photochemistry,crutzen1970influence,stolarski1974stratospheric,wofsy1975chemistry} and the cause of the Southern Hemisphere ozone hole \cite{johnston1971reduction,molina1974stratospheric,crutzen1986nitric,toon1986condensation,solomon1986depletion,mcelroy1986reductions,tung1986antarctic,molina1987production}. These pioneering works have significantly extended our understanding of the complex dynamics of stratospheric variations.

One noteworthy observation during this period is that advances in theory have often been promoted by unexplained phenomena. However, none of these phenomena were anticipated by theory \citep{baldwin2019100}. This is different from the development of physics, where theory can also predict phenomena. For example, the discoveries of black holes \citep{ruffini1971introducing}, gravitational waves \citep{castelvecchi2017gravitational}, and the Higgs boson \citep{higgsphysics} were all guided by theoretical predictions. 

This difference arises because the stratosphere, as a subsystem of the Earth's atmosphere, is a complex system \citep{fan2021statistical}. It is governed by intricate atmospheric dynamic processes, which are in turn regulated by atmospheric chemical processes. Waves and their interaction with the mean flow play a crucial role in driving these processes \citep{andrews1978exact,pfeffer1981wave}. Moreover, these waves span a wide range of scales, from small-scale gravity waves to global-scale planetary waves \citep{baldwin2019100}. 
The renowned physicist Philip W. Anderson’s statement ``More is Different'' vividly captures the essence of emergent behavior in complex systems \citep{anderson1972more}. He highlighted that as systems grow in scale and complexity, entirely new properties emerge---properties that cannot be understood by simply extrapolating from the behavior of individual components. Understanding such emergent behaviors requires fundamental research.
Therefore, the stratosphere, as a complex system, demands the study of emergent large-scale climate patterns of its variations, which is as essential and fundamental as understanding the dynamics that govern it.

EMA is an efficient tool used to study collective behaviors, emergence, phase transitions, and evolution in complex systems. This approach has already been widely applied across various complex systems, including physical systems, biological systems, and climate systems \citep{sun2021eigen}. (See \hl{Appendix} \ref{appA})
. For example, it has been used to study equilibrium and non-equilibrium phase transitions \citep{hu2019condensation,li2021discontinuous,zhang2024eigen,hu2023quantum,li2024exploring}; to predict extreme El Ni\~{n}o events \citep{ma2024increased}; to investigate vegetation growth and evolution patterns in the Heihe River Basin \citep{wang2024holistic}; to analyze the characteristics of stratospheric ozone distribution \citep{chen2021eigen}; and to explore dominant patterns of brain activities \citep{chen2023leading}.
The concept behind the EMA originates from the statistical physics of Gibbs, in which the states of all particles in a system at any given time are considered a microstate, and the microstates over a period of time are treated as an ensemble of the complex system \citep{sun2021eigen}. By calculating the eigenvectors of the correlation matrix between these microstates, the system's eigen microstates are obtained, which are uncorrelated with each other. 

The most critical aspect of EMA is the definition of microstates within a system. This requires a detailed analysis based on the characteristics of the system and the problems being addressed, as well as the selection of appropriate variables (details of the EMA are provided in Section \ref{sec:Eigen_Microstate_Approach}). 
Wind is a key variable for characterizing the stratosphere, as the QBO and the polar vortex are defined using wind-based indices. Furthermore, the stratosphere is part of near-Earth space, a transitional region where spacecraft and high-altitude vehicles operate. The development of near-Earth space vehicles is crucial for defense and military applications, while low-power near-Earth space vehicles can be employed for purposes such as meteorological monitoring, atmospheric environmental protection, and ground-based remote sensing. Understanding the variations of stratospheric wind is, therefore, vital for guiding the design and operation of these vehicles.
Given that the zonal wind in the stratosphere is an order of magnitude stronger than the meridional wind, in this paper, we will use the EMA to study the emergent large-scale patterns of global stratospheric zonal winds.
Although large-scale modes in the stratosphere have been studied for decades, it remains meaningful to investigate the modes of stratospheric zonal winds using EMA for two main reasons. First, by defining microstates more effectively through EMA, it becomes possible to uncover new modes that were previously unidentified. Second, the contributions and spatial patterns of existing modes may change over time. 

The remainder of this paper is organized as follows: Section \ref{sec:data_and_methods} introduces the data and methods. Section \ref{sec:results} presents the results, and Section \ref{sec:Conclusions} provides a summary.

\section{Data and Methods}
\label{sec:data_and_methods}
\subsection{Data}
\label{sec:data}
This study utilizes global zonal wind and sea surface temperature data from the ERA5 (ECMWF Reanalysis v5) monthly reanalysis dataset \citep{era5citation}, covering the period from 1980 to 2022, with a total of \(M\) = 43 × 12 = 516 temporal observations. The spatial resolution of ERA5 is 0.25° × 0.25° in latitude and longitude. For this study, we downsampled the data to a 1° × 1° resolution, where the value at each grid point represents the corresponding 1° × 1° cell, resulting in \(N\) = 360 × 181 = 65,160 discrete grid points. ERA5 provides wind speed data at 37 different pressure levels, ranging from the surface (1000 hPa) to the top of the stratosphere (1 hPa). For this study, we selected four levels---70, 50, 30, and \linebreak 10 hPa---covering the lower to middle stratosphere. 
The sudden stratospheric warmings (SSWs) refer to the breakdown of the polar vortex, accompanied by a rapid descent and warming of the air in polar latitudes \citep{baldwin2021sudden}. The zonal--mean zonal wind at 60° latitude is commonly used as an indicator of the strength of the polar vortex and to determine whether an SSW event has occurred. Therefore, we defined the Arctic polar vortex index and Antarctic polar vortex index at each pressure level as the zonal mean zonal wind at 60°N and 60°S for their respective pressure levels \citep{liang2023northern}. To further analyze the data, we removed the seasonal cycle by subtracting the monthly mean values across all years, resulting in the deseasonalized Arctic polar vortex index (NPVI) and deseasonalized Antarctic polar vortex index (SPVI).
We also used the QBO index defined as the average zonal wind between 5°N and 5°S \citep{yamazaki2020tropospheric,luo2023key}.
Additionally, we used the Ni\~{n}o 3.4 index provided by NOAA (National Oceanic and Atmospheric Administration) \citep{Rayner2003}. The Ni\~{n}o 3.4 index represents the average sea surface temperatures (SST) anomalies across the Ni\~{n}o 3.4 region (5°N\hl{--} 
5°S, 170°W\hl{--}120°W).

\subsection{Eigen Microstate Approach}
\label{sec:Eigen_Microstate_Approach}
The EMA has been recently developed for studying collective behaviors, emergence, and phase transition of complex systems \citep{hu2019condensation,sun2021eigen}. Here, we will use it to study the stratospheric zonal wind. We take a certain pressure level in the stratosphere as a complex system, which is divided into \(N=360 \times 181 = 65160\) grid points with a spatial resolution of 1°×1°. With a monthly resolution spanning from 1980 to 2022, we obtained \(M=516\) temporal observations.

Due to the influence of solar radiation, the stratospheric zonal wind exhibits pronounced seasonality. However, our primary interest lies in the behavior of the zonal wind beyond its seasonal variations. Therefore, before applying the EMA, we must first remove the seasonal component from the zonal wind data:

Let \(u^y_i(m)\) denote the zonal wind speed at grid point \(i\) in the \(m\)th month of year \(y\), where \(i \in [0,N)\), \(m \in [0,12)\), and \(y \in [0,N_y)\), with \( N_y \) representing the number of years.
Let \(\bar{u}^m_i = \frac{1}{N_y} \sum_{y=0}^{N_y-1} u^y_i(m)\) represent the average zonal wind at grid point \(i\) in the \(m\)th month.

To deseasonalize the zonal wind, we calculate the \hl{following:} 

\[
\tilde{u}^y_i(m) =  u^y_i(m) - \bar{u}^m_i
\]
where \(\tilde{u}^y_i(m)\) represents the deseasonalized zonal wind at grid point \(i\), indicating fluctuations relative to the seasonal trend.

Let \(t = y \times 12 + m\), then the deseasonalized zonal wind \(\tilde{u}^y_i(m)\) at grid point \(i\) can be written as \(\tilde{u}_i(t)\), where \(t \in [0,M)\).
Let \(std(\tilde{u}_i) = \sqrt{\frac{1}{M} \sum_{t=0}^{M-1}[\tilde{u}_i(t)]^2}\) represent the standard deviation of the deseasonalized zonal wind at grid point \(i\).

Due to the spatial heterogeneity of zonal winds at different latitudes and longitudes in the Earth system, the fluctuation amplitude varies across grid points. Therefore, it is necessary to make the fluctuations at each grid point dimensionless by dividing them by their standard deviations. Consequently, we can define the system's microstates as \hl{follows:} 
\[
\mathbf{S(t)}= \begin{bmatrix} \frac{\tilde{u}_0(t)}{std(\tilde{u}_0)}  \\  \dots \\ \frac{\tilde{u}_i(t)}{std(\tilde{u}_i)} \\ \dots \\
\frac{\tilde{u}_{N-1}(t)}{std(\tilde{u}_{N-1})} \\ \end{bmatrix}
\]

Taking all microstates at various times, we obtain an \( N \times M\) ensemble matrix \(\mathbf{A}\) with the following elements: \({A}_{i}(t)=\frac{1}{\sqrt{C_{0}}} {S}_{i}(t)\), where \(\sqrt{C_0}= \sum_{i=0}^{N-1}\sum_{t=0}^{M-1}{S}_i^2(t)\).

The spatial correlation between grid points can be calculated as \(K_{ij} = \sum_{t=0}^{M-1}A_i(t)A_j(t)\), resulting in the spatial correlation matrix \(\mathbf{K}=\mathbf{A}\mathbf{A}^T\). The eigenvectors of the matrix \(\mathbf{K}\) form a unitary matrix \(\mathbf{U}=\begin{bmatrix} \mathbf{U_1}, \mathbf{U_2},..., \mathbf{U_N}\end{bmatrix}\). Similarly, the temporal correlation between microstates can be calculated as \(C_{tt'} = \sum_{i=0}^{N-1}A_i(t)A_i(t')\), resulting in the temporal correlation matrix \(\mathbf{C}=\mathbf{A}^T\mathbf{A}\). The eigenvectors of the matrix \(\mathbf{C}\) form a unitary matrix \(\mathbf{V}=\begin{bmatrix} \mathbf{V_1}, \mathbf{V_2},..., \mathbf{V_M}\end{bmatrix}\).

\hl{Mathematically, }
 the eigendecomposition of the correlation matrices is equivalent to performing singular value decomposition (SVD) on the ensemble matrix \(\mathbf{A} = \mathbf{U} \cdot \mathbf{{\Sigma}} \cdot \mathbf{V}^T \), where \(\mathbf{{\Sigma}} = \text{diag}(\sigma_1, \sigma_2, ..., \sigma_R)\), and \(R = \text{min}(N, M)\) is the rank of the ensemble matrix, with \(\sigma_I (I=1,...,R)\) being the singular values arranged in descending order.
\(\mathbf{U}\) and \(\mathbf{V}\) are unitary matrices composed of the eigenvectors corresponding to the spatial correlation matrix \(\mathbf{K}\) and the temporal correlation matrix \(\mathbf{C}\), respectively. Based on the SVD, the ensemble matrix \(\mathbf{A}\) can be expressed as follows:
\[
\mathbf{A}=\sum_{I=1}^{R}\sigma_I \mathbf{A}_I^e
\]
where \(\mathbf{A}_I^e=\mathbf{U}_I \otimes  \mathbf{V}_I\), with \((A_I^e)_{it} = U_{iI}V_{tI}\). The EMA decomposes the spatiotemporal evolution of a complex system into summations of eigenmodes with different weights, where \(\mathbf{U}_I\) and \(\mathbf{V}_I\) represent the spatial pattern and temporal evolution of the \(I\)th eigen microstate (EM), respectively.
Since the definition of the ensemble matrix ensures that \(\sum_{I=1}^{R} \sigma_I^2 = 1\), the contribution of each eigen microstate can be represented by \(W_I=\sigma_I^2\). A larger contribution indicates that the EM contributes more to the original ensemble matrix.

Sometimes, the fluctuation amplitudes of the system may be affected by seasonality, which is reflected in the temporal evolutions as different amplitudes of \(\mathbf{V}_I(t)\) in different months. To describe the relationship between the fluctuation amplitudes and the month, we can define the seasonal variance of the mode in the \(m\)th month as follows:
\[
Var_{m}=\sum_{t \in {\mathbb{S}_m} }V_I(t)^2
\]
where \(\mathbb{S}_m\) is the set of dates that belong to month \(m\). Since \(\mathbf{V}_I\) is a unit vector, we can deduce that \(\sum_{m=1}^{12}Var_{m}=1\).

Reversing the signs of both the spatial pattern \(\mathbf{U}_{I}\) and temporal evolution \(\mathbf{V}_{I}\) (i.e., multiplying by \(-1\)) simultaneously does not affect the results. For ease of comparison, all eigen microstate results in this paper have been adjusted to ensure that the directions of \(\mathbf{U}_{I}\) and \(\mathbf{V}_{I}\) obtained at different pressure levels are aligned.

\subsection{Temporal Correlation Coefficient}
The temporal correlation coefficient measures the degree of linear correlation between two time series, \(X\) and \(Y\). It is defined as follows:
\[ C_{XY} = \frac{ \frac{1}{M} \sum_{t=0}^{M-1} (X_t - \bar{X})(Y_t - \bar{Y})}{\sqrt{ \frac{1}{M} \sum_{t=0}^{M-1} (X_t - \bar{X})^2 \frac{1}{M} \sum_{t=0}^{M-1} (Y_t - \bar{Y})^2}}  = \frac{ \sum_{t=0}^{M-1} (X_t - \bar{X})(Y_t - \bar{Y})}{\sqrt{ \sum_{t=0}^{M-1} (X_t - \bar{X})^2  \sum_{t=0}^{M-1} (Y_t - \bar{Y})^2}}
\]
\(X_t\) and \(Y_t\) represent the values of time series \(X\) and \(Y\) at time \(t\), respectively; \(\bar{X}\) and \(\bar{Y}\) denote the mean values of time series \(X\) and \(Y\), respectively.

The value of \(C_{XY}\) ranges from $-$1 to 1, where a coefficient of 1 signifies perfectly identical linear trends between the series, while -1 indicates completely opposite trends.

In certain scenarios, time-lagged correlation coefficients are utilized to evaluate predictive performance between time series \(X\) and \(Y\). Assuming \(X\) leads \(Y\) (\(\tau>0\)), the time-lagged correlation coefficients are defined as follows:
\[
C_{XY}(\tau) = \frac{ \frac{1}{M-\tau} \sum_{t=\tau}^{M-1} (X_{t-\tau} - \bar{X})(Y_t - \bar{Y})}{\sqrt{ \frac{1}{M-\tau} \sum_{t=\tau}^{M-1} (X_{t-\tau} - \bar{X})^2 \frac{1}{M-\tau} \sum_{t=\tau}^{M-1} (Y_t - \bar{Y})^2}} 
\]
\[
\begin{aligned}
&= \frac{\sum_{t=\tau}^{M-1} (X_{t-\tau} - \bar{X})(Y_t - \bar{Y})}{\sqrt{\sum_{t=\tau}^{M-1} (X_{t-\tau} - \bar{X})^2 \sum_{t=\tau}^{M-1} (Y_t - \bar{Y})^2}}
\end{aligned}
\]

To assess the statistical significance of the correlation coefficient, a significance test is often required. Traditional significance tests for correlation coefficients assume that the data are independent and follow a Gaussian distribution \citep{ross2017introductory}. However, this assumption is clearly unsuitable for climate time series due to their autocorrelations and inherent distributions. Therefore, it is necessary to design a null model tailored to the characteristics of climate time series for significance testing \citep{guez2014influence}.

For instance, when calculating the correlation coefficient between EM2 and the global sea surface temperature (SST) series (in Section \ref{sec:results_EM2}), we generated synthetic surrogate data from the EM2 and SST series individually using the Amplitude adjusted Fourier transform (AAFT) method \citep{lancaster2018surrogate,theiler1992testing} and computed the correlation coefficients 1000 times. The AAFT surrogate not only preserves the distribution of the series but also maintains its power spectra. The 90th and 10th percentiles of these 1000 samples were then used as the positive and negative thresholds for the correlation coefficients, respectively. Correlation coefficients exceeding these thresholds are considered statistically significant.

\subsection{Linear Prediction Model}
\label{sec:Linear_Prediction_Model}
To quantitatively evaluate the predictive capability of EM4 for the Antarctic polar vortex in Section \ref{sec:EM4}, we designed a linear prediction model. This model uses the temporal evolution sequence of EM4 to predict the (SPVI) (see Section \ref{sec:data}). The prediction model is expressed as follows:

\[ SPVI(t) = a \times EM4(t-2) + b + R(t) \]

\textls[-25]{Here, \(a\) and \(b\) are parameters learned through training, and \(R(t)\) represents the residual.}

We trained the model using only data prior to the prediction time. For example, when predicting SPVI at time \(t\), data from time 0 to \(t-1\) were used as the training set. The training process involved deriving the eigen microstates from the training set, identifying the fourth mode and its temporal evolution, and utilizing the SPVI prior to time \(t\) with EM4's temporal evolution to learn parameters \(a\) and \(b\). Then the trained model, along with the value of EM4's temporal evolution at \(t-2\), was used to predict SPVI at time \(t\). The reason for using a two-month lag in this prediction model will be detailed below in \linebreak Section \ref{sec:EM4}).

\subsection{Spectral Peak Significance Test}
\label{sec:SPD_Test}
The spectral significance test is used to assess whether a detected spectral peak in the power spectrum of the signal is statistically significant or could solely originate from (correlated) random noise. 

The significance of a spectral peak is tested by comparing it against a red noise spectrum of a first-order autoregressive process fitted to the data. To assess the statistical significance of a spectral peak, we focus on the ratio of variances between the power spectrum of the signal, denoted as \( \phi_1 \), and the corresponding red noise power spectrum \( \phi_0 \), which is fitted from data. This ratio can be tested using the F-statistic:

\[
F = \frac{S_1^2}{S_0^2}
\]

where \( S_1^2 \) and \( S_0^2 \) are the variances of the signal spectrum and the red noise \linebreak spectrum, respectively. 

For the signal \( y_{1}(t) \), its autocorrelation function, \( \gamma_{1}(\tau) \), and its power spectrum, \( \phi_{1}(\omega) \), are Fourier transforms of each other. For red noise, we use the discrete red noise spectrum developed by Gilman et al. (1963) \citep{Gilman1963} where its unnormalized form is given by \mbox{the following:} 
\[
\phi_{0}(\omega) = \frac{1 - \rho^2}{1 - 2 \rho \cos \left( \frac{h \pi}{N/2} \right) + \rho^2}
\]

where \( h = 0,1,2,\dots,N/2 \), N is the length of the time series and \( \rho \) is the lag-1 autocorrelation of the time series and we use the lag-1 autocorrelation from the signal \( y_{1}(t) \) as \linebreak its estimate.

The F-statistic requires the degrees of freedom \( \nu_1 \) for the numerator and \( \nu_0 \) for the denominator.
For red noise \( \nu_0 \), it is typically assumed to be a large number.
For the real-time signal \( \nu_1 \), we use \( N / M^* \) as its estimate, where \( N \) is the length of the time series and \( M^* \) is the number of independent spectral estimates. For the discrete power spectrum, \( M^* = N / 2 \), and so \( \nu_1=2 \).

The steps for performing the spectral peak significance test are as follows:
\begin{enumerate}
  \item Calculate the power spectrum of the signal.
  \item Estimate the power spectrum of red noise based on the signal's lag-1 autocorrelation.
  \item Calculate the ratio of the signal's power spectrum to the red noise power spectrum.
  \item Perform a significance test on the ratio using the F-statistic.
\end{enumerate}



\section{Results}
\label{sec:results}

The stratosphere extends approximately from the tropopause at 70 hPa to the stratopause at 0.1 hPa. However, 99\% of the atmospheric mass is concentrated below 10 hPa \citep{mohanakumar2008stratosphere}. In addition, this study focuses on the stratosphere, aiming to minimize the influence of adjacent layers on mass continuity. To reduce the impact of the boundary layers near the troposphere and mesosphere, we did not select the layers close to these boundaries, such as 100 hPa or 1 hPa. Therefore, our study focuses on the following four representative layers: 70, 50, 30, and 10 hPa.

\begin{figure}[H]
  
  \includegraphics[scale=0.6]{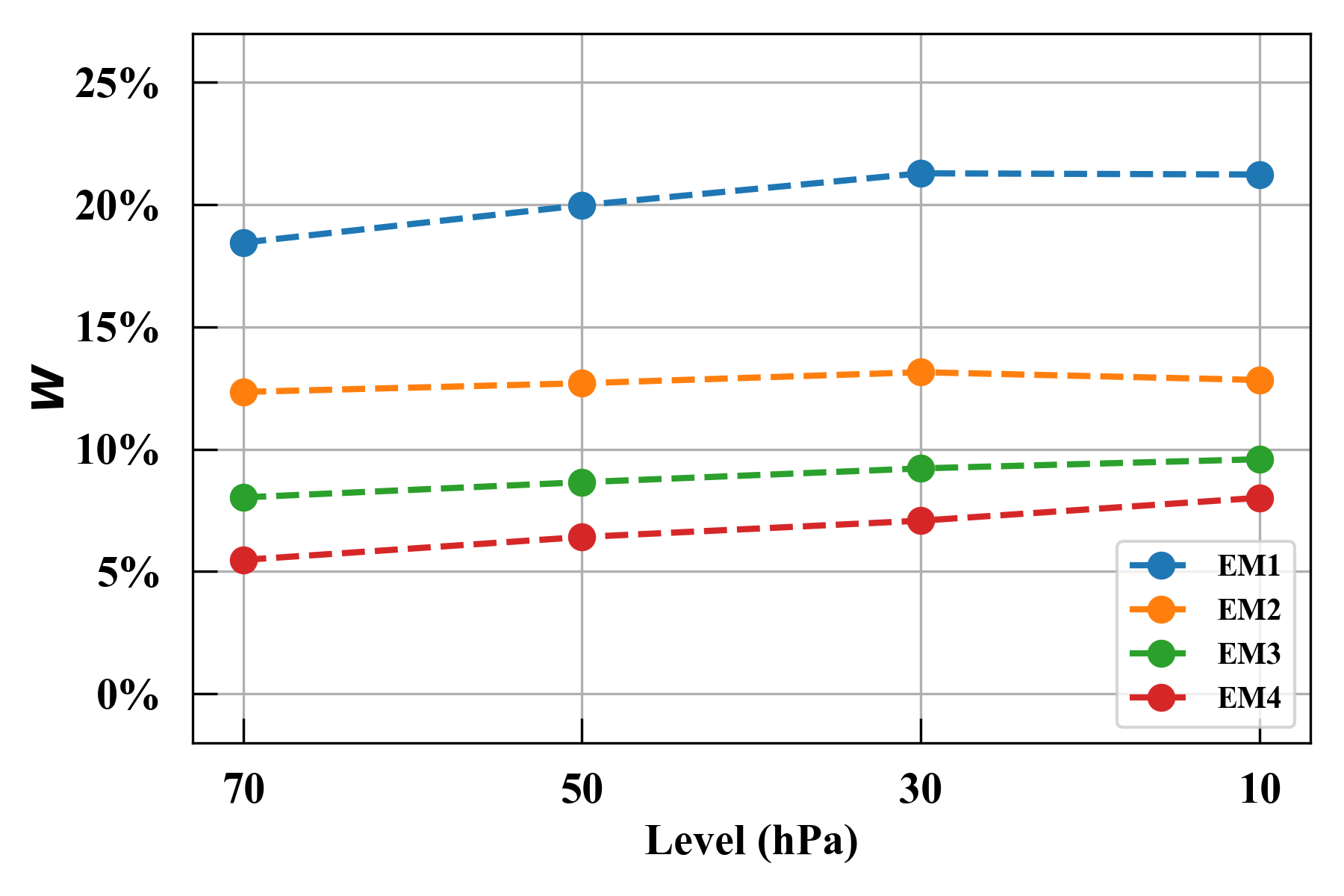}
  \caption{The contributions of the top four eigen microstates (EM1 to EM4) at 70--10 hPa. Blue, orange, green, and red points indicate EM1 to EM4, respectively; derived from ERA5 Reanalysis data (1980--2022; see Section \ref{sec:data_and_methods}).}
  \label{EM_weight}
\end{figure}

Figure \ref{EM_weight} shows the contributions of the top four eigen microstates at 70, 50, 30, and \mbox{10 hPa.} Together, they contribute about 40--50\% of the variability in the stratospheric zonal winds.
The first eigen microstate (EM1) has the highest contribution, around 20\%. The second to fourth eigen microstates (EM2 to EM4) contribute approximately 13\%, 9\%, and 7\% of the variance, respectively.  


\subsection{EM1 and Its Relationship with QBO}



\begin{figure}[H]
  \centering
  \includegraphics[scale=0.67]{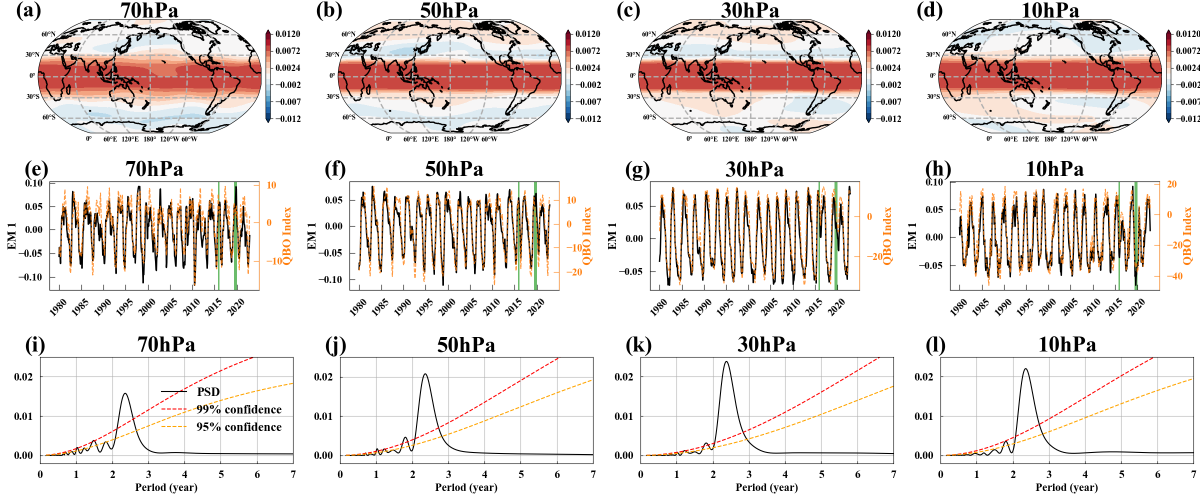}
  \caption{EM1. (\textbf{a}–\textbf{d}) Spatial patterns of EM1 at 70, 50, 30, and 10 hPa, where positive values (red) indicate westerlies and negative values (blue) indicate easterlies; (\textbf{e}–\textbf{h}) temporal evolution of EM1 at 70, 50, 30, and 10 hPa (black solid lines) with QBO index (orange dashed lines), showing high correlation (0.88, 0.94, 0.96, 0.95). Green-shaded areas indicate moments when QBO disruptions occurred. (\textbf{i}–\textbf{l}) Power spectral density of EM1's temporal evolution at 70, 50, 30, and 10 hPa, with red and orange dashed lines representing 99\% and 95\% confidence levels.}
  \label{EM1}
\end{figure}

Figure \ref{EM1}a–d show EM1's spatial patterns at four different pressure levels. \linebreak Figure \ref{EM1}e–h depict the corresponding temporal evolutions (black solid lines), while \linebreak Figure \ref{EM1}i–l present their power spectral densities. EM1’s spatial patterns across all pressure levels exhibit similar characteristics, with a strong correlation in zonal winds within the tropical stratosphere, approximately between 20° N and 20° S. The power spectra of EM1 reveal a significant cycle of around 2.4 years, indicating that the zonal winds in the tropical stratosphere reverse direction approximately every 28 months. 

The QBO is a quasi-periodic oscillation phenomenon in the tropical stratosphere, where easterlies and westerlies alternately reverse, with an average cycle of about \mbox{28 months.} These alternating easterlies and westerlies descend downward at a rate of about 1 km per month \citep{baldwin2001quasi}. Thus, it can be inferred that EM1 is a mode associated with the QBO.
The QBO was discovered in the early 1960s through observations of stratospheric zonal winds at Canton Island, which revealed a 2-year round oscillation \citep{ebdon1960notes,reed1961evidence}. Subsequently, theories were developed to understand this phenomenon \citep{lindzen1968theory,holton1972updated}, explaining that the QBO is a wave-driven circulation. The driving waves include small-scale gravity waves, global-scale Kelvin waves, Rossby waves, as well as mixed Rossby and gravity \linebreak waves \citep{kawatani2010roles,richter2014simulation,kim2015momentum,kim2015contributions,pahlavan2021revisiting,ern2014interaction}. Although the mechanisms of QBO are known, the relative importance of these driving waves remains uncertain. State-of-the-art numerical models are still unable to represent QBO precisely, particularly during QBO disruption events \citep{newman2016anomalous,saunders2020quasi}. 

The first QBO disruption occurred in 2015–2016. An anomalous upward displacement of westerly winds from ~30 hPa to 15 hPa interrupted the normal downward propagation of the easterly phase. This disruption was unprecedented since observations began in 1953 \citep{newman2016anomalous}. Coincidentally, a similar QBO disruption occurred again four years later, during the winter of 2019--2020 \citep{saunders2020quasi}. Many studies have analyzed the causes of these two QBO disruptions, but the mechanisms behind their occurrence remain a topic of debate \citep{wang2023revisit,osprey2016unexpected,kang2022role}. Some research suggests that, against the backdrop of global warming, QBO disruptions may become increasingly frequent \citep{anstey2021prospect}.

The importance of the QBO is self-evident due to its wide-ranging impacts. For instance, the QBO can influence the strength of the polar vortex, as described by the Holton--Tan relationship \citep{holton1980influence,anstey2014high,garfinkel2018extratropical,rao2020impact}.
Randel et al. used singular-value decomposition and regression analyses to identify QBO signals in ozone and nitrogen dioxide data in the stratosphere \citep{randel1996isolation}. Rao and Yu et al. explored the climatology and trends of the northern winter stratospheric residual mean meridional circulation, which are influenced by the QBO, ENSO, and solar cycles \citep{rao2019evaluating}.
Moreover, the QBO can also impact jet streams \citep{garfinkel2011influence,wang2018interannual} and the Madden--Julian oscillation (MJO) \citep{martin2021influence}. A recent review article summarized four pathways of QBO teleconnections, providing a comprehensive overview of its broader impacts \citep{anstey2022impacts}.

Currently, the QBO index is primarily used to represent the state and phase of the QBO and to measure its impact by correlating the index with climate variables in other regions or through composite analyses \citep{luo2023key,yamazaki2020tropospheric,zhang2024stratospheric,wang2019impact}. 

Selecting an appropriate QBO index is crucial for obtaining reliable research results. The most widely used QBO index is based on radiosonde observations collected in Singapore (1° N, 104° E) and compiled by the Free University of Berlin \citep{naujokat1986update}. 
However, observations from a single meteorological station are insufficient to fully capture the QBO phenomenon due to its regional heterogeneity. The strength of the QBO signal diminishes as latitude increases \citep{dunkerton1985climatology}. A QBO index constructed from a larger set of data can better reflect the overall characteristics of the QBO. With advancements in observations and climate modeling, stratospheric reanalysis data have become increasingly accurate, enabling the construction of more precise QBO indices. Some studies have defined the QBO index using the average zonal wind between 5° N and 5° S \citep{yamazaki2020tropospheric,luo2023key}. Considering the descending nature of the QBO, the phase at different pressure levels is also an important feature. Other studies have employed EOF (empirical orthogonal function)-based methods to measure both the amplitude and phase properties of the QBO \citep{xu2023ceof}. 

We compared EM1’s temporal evolutions with the commonly used QBO index based on zonal wind averages from 5° N to 5° S (the orange dashed line in Figure \ref{EM1}e–h). The two align well, with a correlation coefficient exceeding 0.88. However, slight differences are observed during the QBO's maximum easterly and westerly phases, as well as during periods of QBO disruption (green-shaded areas). 
Notably, EM1's spatial patterns \linebreak (Figure \ref{EM1}a–d) indicate that the zonal extent of the QBO is broader than 5° N to 5° S. This is because the EMA captures the common features (emergent characteristics) across all spatial grid points. By normalizing each grid point by its own standard deviation, the heterogeneity between spatial grid points is reduced, enabling the detection of even very weak QBO signals at higher latitudes.
This suggests that EM1’s temporal evolutions, derived from global zonal wind using the EMA, provide more detailed information about the QBO's distribution. Additionally, EM1's spatial patterns clearly illustrate the range and intensity distribution of the QBO phenomenon.

\subsection{EM2 and The Arctic Polar Vortex Variations}
\label{sec:results_EM2}



The stratospheric polar vortex is a large-scale cyclonic circulation located in the stratosphere over the polar regions. It forms in autumn as temperatures in the polar stratosphere drop and weaken or break down in spring \citep{waugh2017polar}. The polar vortices play a critical role in stratospheric and tropospheric weather patterns, including their influence on the ozone layer \citep{zuev2021influence,schoeberl1991dynamics} and sudden stratospheric warmings (SSWs) \citep{roy2022dynamical,waugh2017polar,thompson2002stratospheric,lim2019australian,kolstad2010association,baldwin2001stratospheric,baldwin2021sudden,overland2020polar}.


SSWs are some of the causes of extreme cold events, which have a significant impact on human life. Extensive research has been conducted on the causes, impacts, and predictability of SSWs. For instance, Yu and Cai revealed the linkage between SSWs and surface cold-air outbreaks (CAOs) \citep{yu2018linkage}. Rao et al. demonstrated the deterministic predictable limit of the February 2018 major SSW using a climate system model \citep{rao2018stratospheric}. Bao et al. classified the tropospheric precursor patterns of SSW \citep{bao2017classifying}, while Ma et al. analyzed two possible causes of the rare Antarctic SSW in 2019 \citep{ma2022possible}.

\begin{figure}[H]
  \centering
  \includegraphics[scale=0.67]{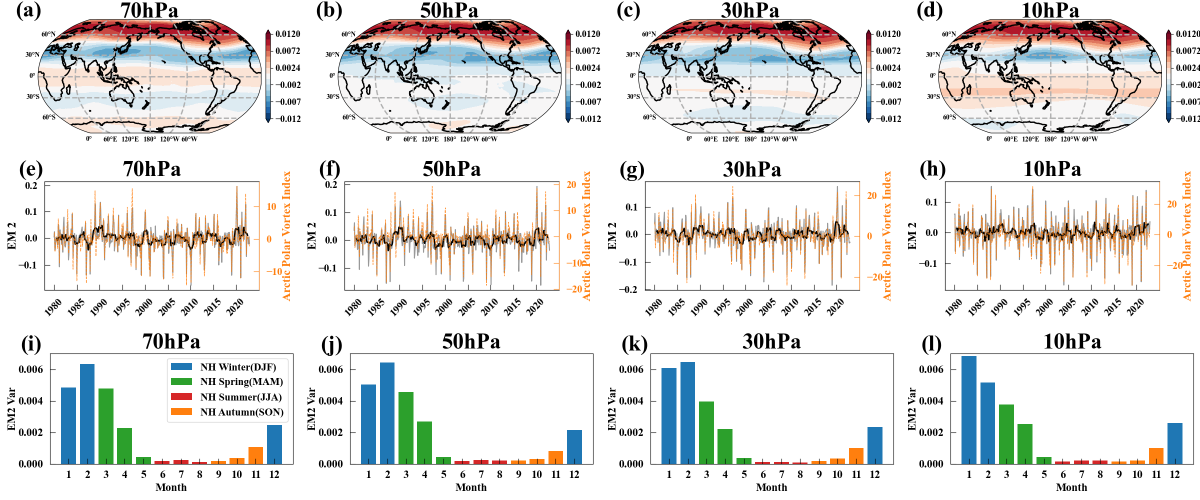}
  \caption{EM2. (\textbf{a}–\textbf{d}) Spatial patterns of EM2 at 70, 50, 30, and 10 hPa, with positive values (red) indicating westerlies and negative values (blue) indicating easterlies. (\textbf{e}–\textbf{h}) Temporal evolution of EM2 at 70, 50, 30, and 10 hPa (gray solid lines), with the orange dashed lines representing the deseasonalized Arctic polar vortex index (NPVI) (see Section \ref{sec:data_and_methods}), showing a high degree of alignment, with correlation coefficients of 0.93, 0.92, 0.93, and 0.96, respectively. The black solid line represents the 12-month smoothed temporal evolution of EM2. Panels (\textbf{i}–\textbf{l}) seasonal variance of EM2 at 70, 50, 30, and 10 hPa, with green, red, orange, and blue representing spring, summer, autumn, and winter in the Northern Hemisphere, respectively.}
  \label{EM2}
\end{figure}

Figure \ref{EM2}a–d show EM2’s spatial patterns at four different pressure levels. They exhibit similar characteristics at each pressure level, primarily concentrated in the mid-to-high latitudes of the Northern Hemisphere. Specifically, strong westerlies are present in the 60° N–90° N region, while easterlies dominate the 30° N–50° N region. In Figure \ref{EM2}e–h, the gray solid lines represent EM2’s temporal evolutions, and the orange dashed lines represent the deseasonalized Arctic polar vortex index (NPVI). The significant overlap of these lines, with correlation coefficients exceeding 0.92, indicates that EM2 represents a mode describing variations of the Arctic polar vortex.

EM2's temporal evolutions exhibit many `spikes', suggesting that the intensity of the Arctic polar vortex fluctuates sharply. To explore the relationship between the severity of these fluctuations and the months of the year, we introduced the seasonal variance metric (see Section \ref{sec:Eigen_Microstate_Approach}) from the EMA, with the results shown in Figure \ref{EM2}i–l. The figure shows that the strongest fluctuations in the Arctic polar vortex occur during the winter and early spring (January--March), which is when the strongest fluctuations in the Arctic polar vortex occur. In contrast, fluctuations are minimal in summer, as the Arctic polar vortex has already collapsed by that time \citep{wei2007dynamical}.

\begin{figure}[H]
  \centering
  \includegraphics[scale=0.67]{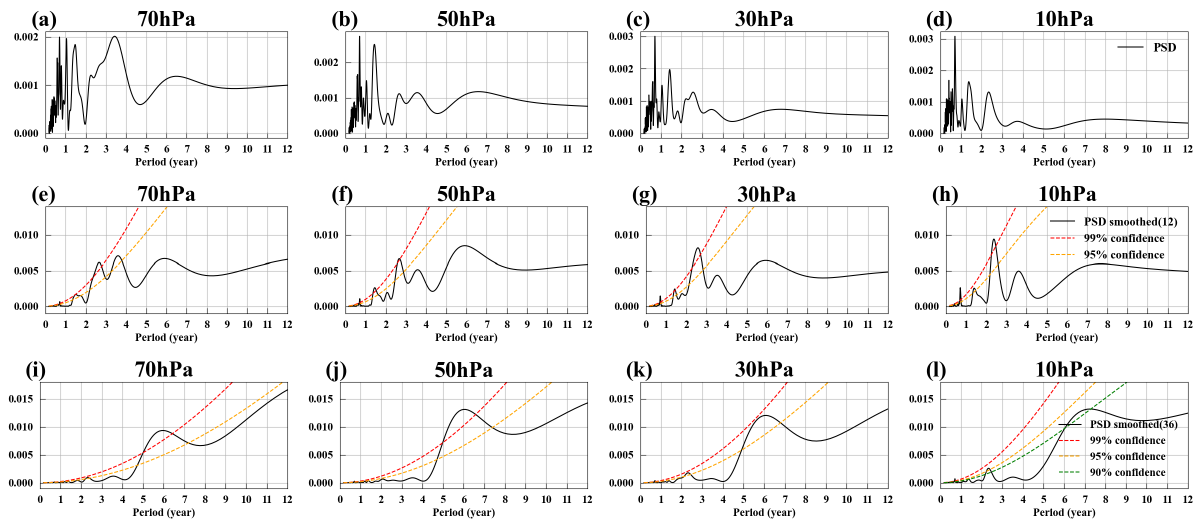}
  \caption{Power spectral density of EM2. (\textbf{a}–\textbf{d}) Power spectral density of EM2 at 70, 50, 30, and \linebreak 10 hPa. (\textbf{e}–\textbf{h}) Power spectral density of EM2 with a 12-month smoothing at 70, 50, 30, and 10 hPa. (\textbf{i}–\textbf{l}) Same as (\textbf{e}–\textbf{h}) but with a 36-month smoothing. Red and orange dashed lines represent 99\% and 95\% confidence levels, respectively; green-dashed line represents 90\% confidence level.}
  \label{EM2PSD}
\end{figure}

Figure \ref{EM2PSD}a–d show the power spectral density of EM2, revealing that EM2 contains not only strong high-frequency signals on an annual scale but also low-frequency trends. By applying a 12-month moving average to EM2, the high-frequency signals are filtered out, resulting in the black solid lines shown in Figure \ref{EM2}e–h. Figure \ref{EM2PSD}e–h display the power spectral density of these smoothed signals. After filtering out the high-frequency signals, EM2 exhibits clear low-frequency trends, with significant cycles around 1.5 years and 2.4–2.6 years in its power spectrum.

\begin{figure}[H]
  \centering
  \includegraphics[scale=0.67]{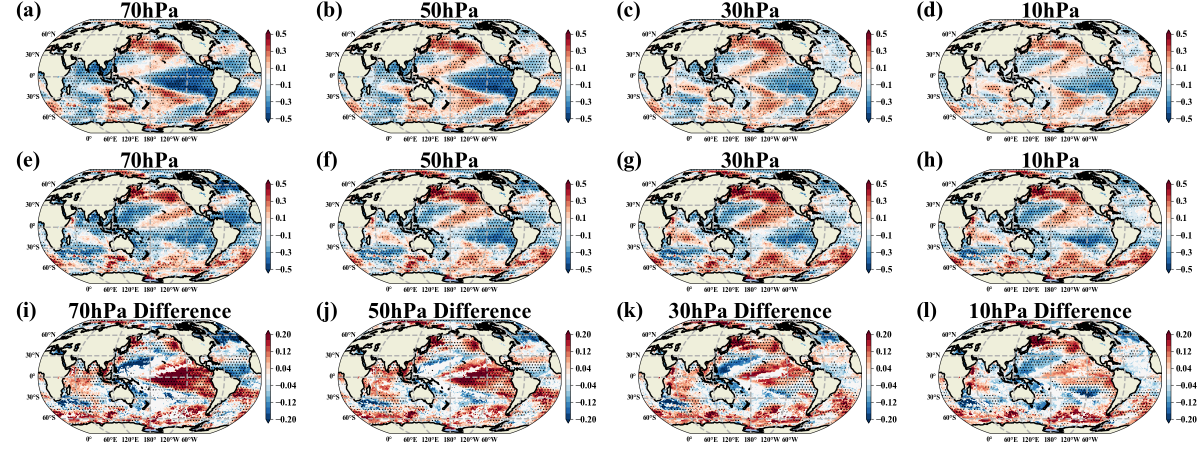}
  \caption{Correlation between EM2 and SST. (\textbf{a}–\textbf{d}) Correlation coefficients between EM2 temporal evolutions at 70, 50, 30, and 10 hPa and SST (both time series are smoothed over 12 months before the correlation analysis). (\textbf{e}–\textbf{h}) Same as (\textbf{a}–\textbf{d}), but with 36-month smoothing. The dashed area indicates regions where the correlation coefficients exceed the 90th percentile threshold. (\textbf{i}--\textbf{l}) Differences between the 36-month-smoothed correlation and the 12-month-smoothed correlation. Positive values (red) in the figure indicate that the 36-month-smoothed correlation is higher than the 12-month-smoothed correlation. Only regions where the correlations from both smoothing methods share the same sign are shown in the figure; regions with opposite signs are set to 0. The dotted areas indicate regions that are statistically significant in both methods.}
  \label{EM2_SST_Corr}
\end{figure}

We know that the 2.4–2.6-year cycle originates from the influence of the QBO. The QBO affects the Arctic polar vortex through the Holton--Tan Effect, where the polar vortex in the Northern Hemisphere's winter is strengthened (weakened) when the QBO at \linebreak 50 hPa is in its westerly (easterly) phase \citep{holton1980influence,lu2014mechanisms}. 
But the origin of the approximately 1.5-year cycle remains unclear. Given that stratospheric circulation is driven by wave-driven residual circulation, and that sea surface temperatures significantly impact these waves, we hypothesize that the 1.5-year cycle of the Arctic polar vortex originates from oceanic influences. To investigate this, we first applied a 12-month smoothing to EM2’s temporal evolutions to remove high-frequency signals and then calculated the correlation coefficients with global SST. Since the influence of SST on the Arctic polar vortex may be delayed, we calculated the correlation coefficients by leading the SST time series by 0–4 months ahead of EM2, selecting the maximum absolute correlation coefficient to generate the correlation map between SST and EM2. The results are shown in Figure \ref{EM2_SST_Corr}a–d. It can be seen that SST in the central and western Pacific are significantly negatively correlated with EM2, especially at 70 hPa and 50 hPa. The distribution of correlation coefficients exhibits the characteristic horseshoe pattern associated with ENSO-related sea surface temperatures.

\begin{figure}[H]
  \centering
  \includegraphics[scale=0.67]{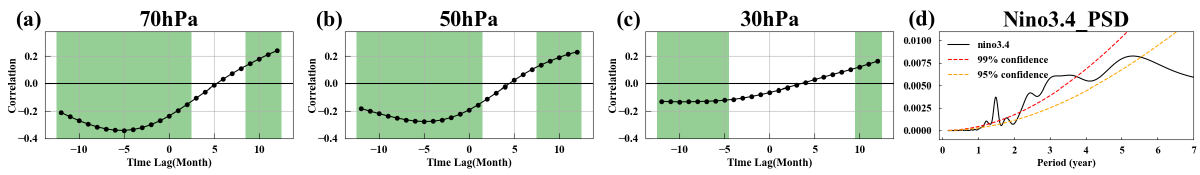}
  \caption{Lead--lag correlation between EM2’s temporal evolutions and the Ni\~{n}o 3.4 index and power spectral density of the Ni\~{n}o 3.4 Index. (\textbf{a}–\textbf{c}) Lead--lag correlation between EM2 at 70, 50, and \linebreak 30 hPa and the Ni\~{n}o 3.4 index (both times series are smoothed over 12 months before the correlation analysis); green-shaded areas indicate \emph{p} $<$ 0.01. Lag $<$ 0 indicates the Ni\~{n}o 3.4 index leads EM2. \linebreak (\textbf{d}) Power spectral density of the Ni\~{n}o 3.4 index, with red and orange dashed lines representing 99\% and 95\% confidence levels, respectively.}
  \label{EM2_Nino34_Corr}
\end{figure}

The sea surface temperatures in the central and western Pacific are related to ENSO, the most well-known climate phenomenon globally, which has widespread impacts on the world’s climate \citep{yang2018Nino}. Research has shown that during El Ni\~{n}o events, the Brewer--Dobson circulation strengthens, which enhances the upward propagation of planetary waves and subsequently weakens the polar vortex in both hemispheres, while the opposite occurs during La Ni\~{n}a events \citep{domeisen2019teleconnection}. Figure \ref{EM2_Nino34_Corr}a--c show the lead--lag correlation between EM2 and the Ni\~{n}o 3.4 index, indicating a significant negative correlation, with ENSO leading EM2 by about four months.

By using Monte Carlo singular spectrum analysis (MC-SSA), Jevrejeva et al. \citep{jevrejeva2004oceanic} revealed that the periodic signals within ENSO can influence Northern Hemisphere winter climate variability, with a phase lag of approximately three months. Jevrejeva et al. speculated that these signals are most likely transmitted via the stratosphere, with the Arctic oscillation (AO) mediating their propagation through the coupled stratospheric and tropospheric circulation variability, which facilitates vertical planetary wave propagation. This may account for the observed four-month lag in the correlation between ENSO and EM2. Additionally, the power spectrum of the Ni\~{n}o 3.4 index (Figure \ref{EM2_Nino34_Corr}d) also shows a significant 1.5-year cycle. Therefore, the 1.5-year cycle observed in the power spectrum of EM2 may be attributed to the influence of ENSO.

In addition to the 1.5-year cycle and the QBO cycle, the power spectrum of EM2 also shows a significant cycle of approximately 5-6 years at all pressure levels after applying a 36-month smoothing, as shown in Figure \ref{EM2PSD}i–l. Similarly, we applied a 36-month smoothing to both the temporal evolutions of EM2 and SST, then calculated the correlation coefficients, as shown in Figure \ref{EM2_SST_Corr}e–h. After this smoothing, there was a strong positive correlation between the North Pacific SST and EM2. Numerous studies have examined the interaction between North Pacific SST and the Arctic polar vortex. Some research studies indicate that an increase in North Pacific SST weakens the Aleutian Low, thereby reducing the upward flux of planetary waves and strengthening the polar vortex \citep{hu2018recent,li2018connection}. Other studies suggest that variations in Arctic stratospheric circulation can also influence North Pacific SST, triggering the Victoria mode, which in turn influences ENSO \citep{xie2016connection}.
However, the 5--6-year cycle characteristics of the Arctic polar vortex have been scarcely mentioned in existing research, making this topic worthy of more in-depth and detailed study.

The observed 5–6-year cycle could potentially be associated with the 11-year solar sunspot cycle. Many natural terrestrial phenomena are influenced by the 11-year solar sunspot cycle, such as volcanic activity. In recent years, a number of studies have been published on the relationship between solar activity and volcanic activity \citep{gray2010solar,komitov2024possible}. The general conclusion from these studies is that volcanic activity is modulated by sunspots and associated space weather processes, with the strongest relationship observed during two phases of the 11-year solar cycle, near the minimum and maximum of sunspot activity. The proposed physical mechanism involves the forcing of electrical current systems between the ionosphere and the upper lithosphere, driven by solar X-rays and UV radiation during solar maxima (including extreme events such as solar flares) and by galactic cosmic rays (GCRs) during solar minima \citep{komitov2022danjon,komitov2023lower,komitov2024possible}. However, this kind of proposed physical linkage still needs to be further studied.

The impact of sunspot maxima and minima on terrestrial activity may potentially give rise to the observed 5–6-year cycle. However, due to the limited availability of volcanic eruption data, as well as the variation in eruption locations and magnitudes, it is difficult to achieve statistical significance within a 5–6-year time frame. Consequently, studies on volcanic eruption cycles often focus on longer periods \citep{khain2009possible,qu2011periodicity}. 

Further research is needed to investigate and verify the relationship between the 5–6-year cycle and solar sunspot activity.

\subsection{EM3 and The Antarctic Polar Vortex Variations}

\begin{figure}[H]
  \centering
  \includegraphics[scale=0.67]{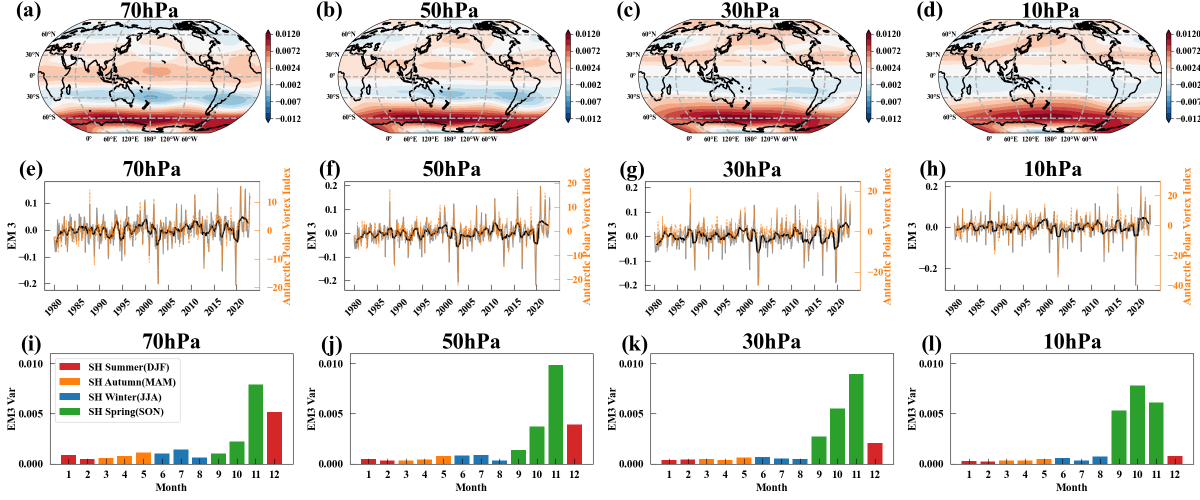}
  \caption{EM3. (\textbf{a}--\textbf{d}) Spatial patterns of EM3 at 70, 50, 30, and 10 hPa, with positive values (red) indicating westerlies and negative values (blue) indicating easterlies. (\textbf{e}--\textbf{h}) Temporal evolutions of EM3 at 70, 50, 30, and 10 hPa (gray solid lines), with the orange dashed lines representing the deseasonalized Antarctic polar vortex index (SPVI) (see Section \ref{sec:data_and_methods}), showing a high degree of alignment, with correlation coefficients of 0.88, 0.93, 0.91, and 0.89, respectively. The black solid line represents the 12-month smoothed temporal evolutions. (\textbf{i}--\textbf{l}) Seasonal Variance of EM3 at 70, 50, 30, and 10 hPa, with green, red, orange, and blue representing spring, summer, autumn, and winter in the Southern Hemisphere, respectively.}
  \label{EM3}
\end{figure}
\begin{figure}[H]
  \centering
  \includegraphics[scale=0.67]{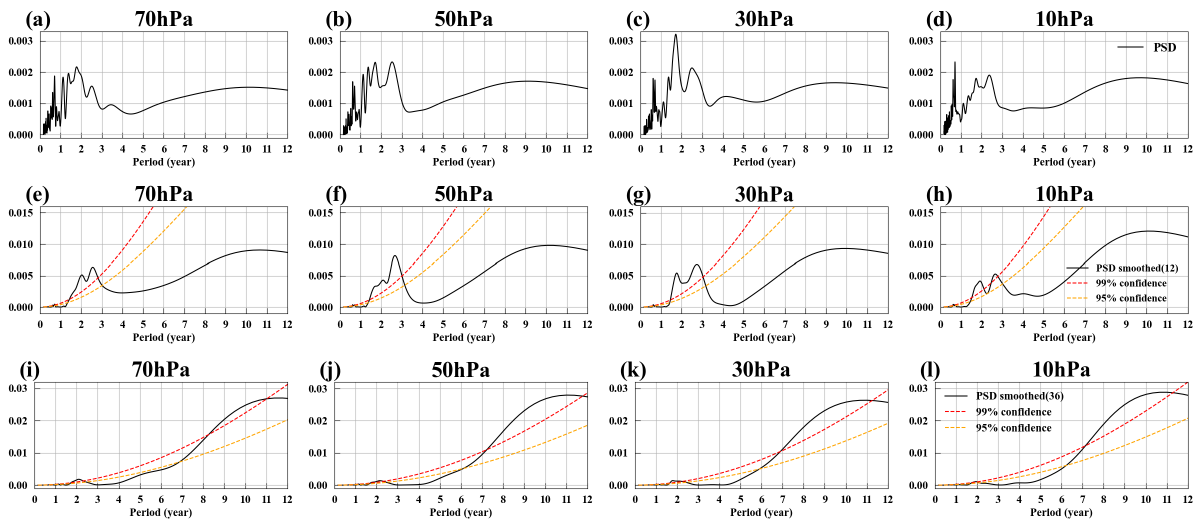}
  \caption{Power spectral density of EM3. (\textbf{a}--\textbf{d}) Power spectral density of EM3 at 70, 50, 30, and \linebreak 10 hPa. (\textbf{e}--\textbf{h}) Power spectral density of EM3 with a 12-month smoothing at 70, 50, 30, and 10 hPa. (\textbf{i}--\textbf{l}) Same as (\textbf{e}--\textbf{h}) but with a 36-month smoothing. Red and orange dashed lines represent 99\% and 95\% confidence levels, respectively.}
  \label{EM3PSD}
\end{figure}

Figure \ref{EM3}a–d show EM3’s spatial patterns at four different pressure levels, with similar characteristics across all levels, primarily featuring strong westerlies near 60° S. In \linebreak Figure \ref{EM3}e–h, the gray solid lines represent EM3’s temporal evolutions, while the orange dashed lines represent the deseasonalized Antarctic polar vortex index (SPVI). These lines overlap significantly, with correlation coefficients exceeding 0.88, indicating that EM3 is a mode reflecting variations in the Antarctic polar vortex. The black solid line represents the 12-month smoothed temporal evolutions, revealing that, in addition to high-frequency variations, EM3 also exhibits low-frequency variation characteristics.
Figure \ref{EM3PSD}e–h show the power spectral density of the 12-month smoothed temporal evolutions, revealing two significant cycles of approximately 1.8–2 years and around 2.6 years. The 1.8–2-year cycle is likely influenced by ENSO, while the 2.6-year cycle continues to be affected by the QBO. Although the impact of the QBO on the Antarctic polar vortex is not as well-known as the Holton--Tan Effect, numerous studies have observed the influence of the QBO on the Southern Hemisphere. For example, the QBO can influence tropical convection, generating Rossby wave trains that propagate southward to high latitudes in the Southern Hemisphere, affecting Antarctic sea ice \citep{yamazaki2021stratospheric}; during the QBO easterly phase, the circumpolar westerlies around Antarctica tend to slow down \citep{rao2023projected}.

After applying a 36-month smoothing to temporal evolutions of EM3, its power spectral density was calculated, as shown in Figure \ref{EM3PSD}i–l. Unlike EM2, EM3 does not exhibit a 5--6-year cycle; instead, it shows a cycle of approximately 11 years. 
Why is the 11-year cycle not significant in the Arctic polar vortex, while it is prominent in the Antarctic polar vortex?  The 11-year cycle may possibly be linked to the 11-year sunspot cycle. \linebreak Figure \ref{EM3}i–l display the seasonal variance of EM3, revealing that the strongest fluctuations in the Antarctic polar vortex occur in the late spring to early summer of the Southern Hemisphere (October to December), which also corresponds to the months with the greatest variability in the Antarctic polar vortex \citep{butler2021wave,zambri2021emergence}. During the polar night of each hemisphere's winter, the polar vortex is shielded from direct solar radiation, making it challenging for the sunspot cycle to exert a direct influence. Moreover, the breakup of the polar vortex in Antarctica occurs later than that of the Arctic \citep{zuev2019cause}. 
However, the physical mechanisms behind this phenomenon require further analysis.

\subsection{EM4: The Southern Hemisphere Dipole Mode}
\label{sec:EM4}
\begin{figure}[H]
  \centering
  \includegraphics[scale=0.67]{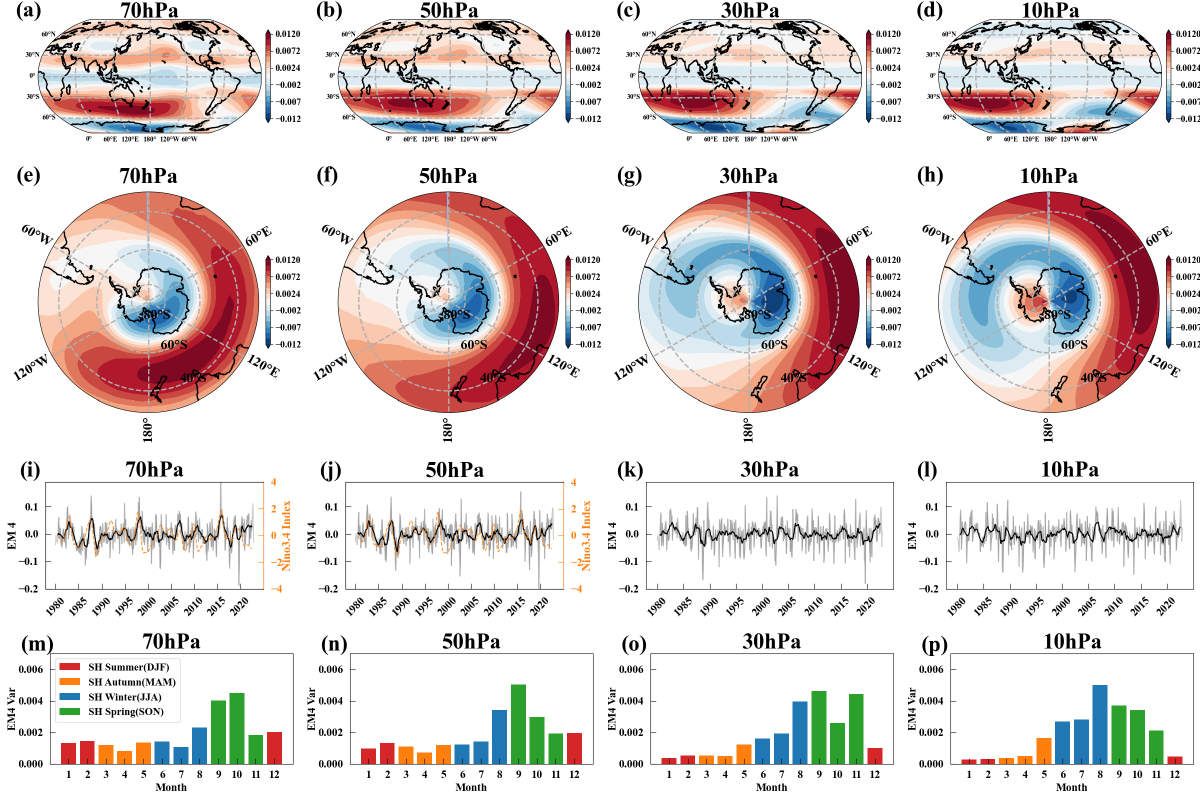}
  \caption{EM4: Southern Hemisphere dipole mode. (\textbf{a}–\textbf{d}) Spatial patterns of EM4 at 70, 50, 30, and 10 hPa, with positive values (red) indicating westerlies and negative values (blue) indicating easterlies. (\textbf{e}–\textbf{h}) the same as (\textbf{a}–\textbf{d}) but in the Antarctic polar projection. (\textbf{i}–\textbf{l}) Temporal evolutions of EM4 at 70, 50, 30, and 10 hPa (gray solid lines), with the black solid line representing the 12-month smoothed temporal evolutions. (\textbf{m}–\textbf{p}) Seasonal Variance of EM4 at 70, 50, 30, and 10 hPa, with green, red, orange, and blue representing spring, summer, autumn, and winter in the Southern \mbox{Hemisphere, respectively.}}
  \label{EM4}
\end{figure}

Figure \ref{EM4}a--d show EM4’s spatial patterns at four different pressure levels, while \linebreak Figure \ref{EM4}e–h present its Antarctic projection. A typical feature of EM4’s spatial patterns is the dipole pattern in zonal winds, divided by the 60° S latitude, with contrasting winds to the north and south of this parallel. As altitude increases, a wave-1 pattern emerges across the eastern and western hemispheres, closely resembling the influence of planetary wave-1 on the Southern Hemisphere. In Figure \ref{EM4}i–l, the gray solid lines represent EM4’s temporal evolutions, while the black solid lines indicate the 12-month smoothed temporal evolutions, which also exhibit low-frequency variations. 
In the mid-to-lower stratosphere (70 and 50 hPa), the 12-month smoothed temporal evolutions show a significant correlation with the Ni\~{n}o 3.4 index with the latter leading by approximately 3 months. The correlation coefficients are 0.51 and 0.50 respectively. The orange dashed lines in Figure \ref{EM4}i,j represent the 12-month smoothed Ni\~{n}o 3.4 index. 

\begin{figure}[H]
  \centering
  \includegraphics[scale=0.67]{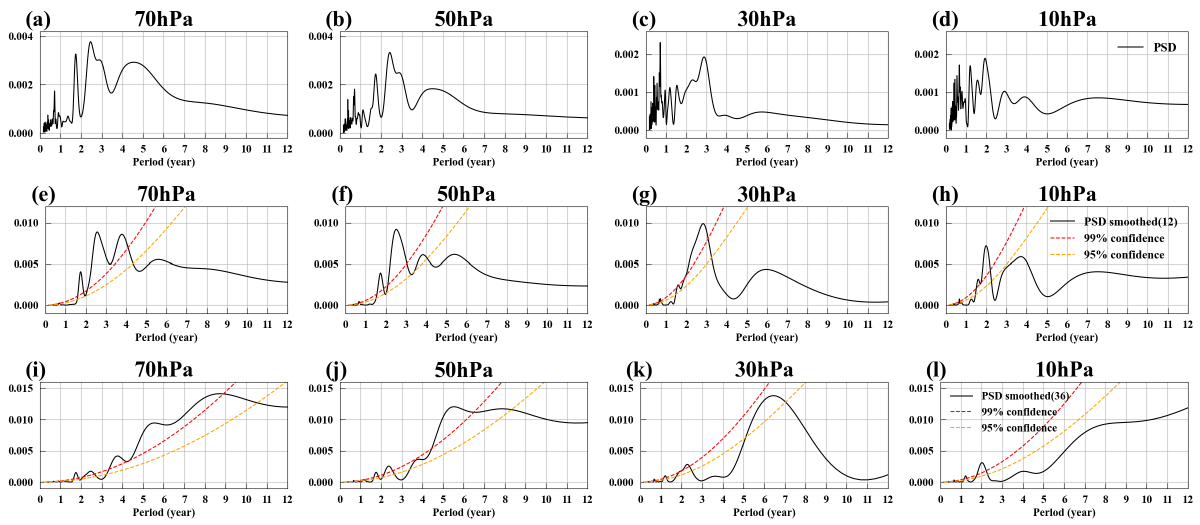}
  \caption{Power spectral density of EM4. (\textbf{a}–\textbf{d}) Power spectral density of EM4 at 70, 50, 30, and \linebreak 10 hPa. (\textbf{e}–\textbf{h}) Power spectral density of EM4 with a 12-month smoothing at 70, 50, 30, and 10 hPa. (\textbf{i}–\textbf{l}) Same as (\textbf{e}–\textbf{h}) but with a 36-month smoothing. Red and orange dashed lines represent 99\% and 95\% confidence levels, respectively.}
  \label{EM4PSD}
\end{figure}

In the mid-to-lower stratosphere (70 and 50 hPa), the power spectral density of EM4 reveals significant cycles of approximately 1.8 years, 2.5–2.6 years, and 3.8 years after a 12-month smoothing (Figure \ref{EM4PSD}e,f), along with a 5.4–5.5 year cycle after a 36-month smoothing (Figure \ref{EM4PSD}i,j). In contrast, these cycles vary in the middle stratosphere (30 and 10 hPa) (Figure \ref{EM4PSD}g,h,k,l).
The seasonal variance of EM4, shown in Figure \ref{EM4}m–p, indicates that the amplitude of EM4 is strongest during the winter and spring in the Southern Hemisphere.

\begin{figure}[H]
  \centering
  \includegraphics[scale=0.67]{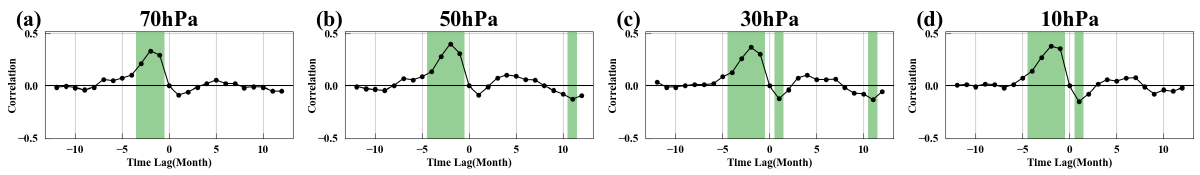}
  \caption{Lead--lag correlation between EM4 and EM3. (\textbf{a}–\textbf{d}) Results at 70, 50, 30, and 10 hPa, respectively. Time Lag $<$ 0 indicates that EM4 leads EM3. The green-shaded areas indicate \emph{p} $<$ 0.01.}
  \label{EM4EM3_Corr}
\end{figure}

We observed an interesting phenomenon: a leading connection between EM4 and EM3. Figure \ref{EM4EM3_Corr}a–d display the lead--lag correlation between EM4 and EM3, where Time Lag $<$ 0 indicates that EM4 leads EM3. The analysis reveals a significant correlation, with a correlation coefficient of around 0.4, when EM4 leads EM3 by two months. As we know, EM3 is a mode that reflects variations in the strength of the Antarctic polar vortex. This suggests that EM4 may have predictive significance for variations in the Antarctic \mbox{polar vortex.}

\begin{figure}[H]
  
  \includegraphics[scale=1.2]{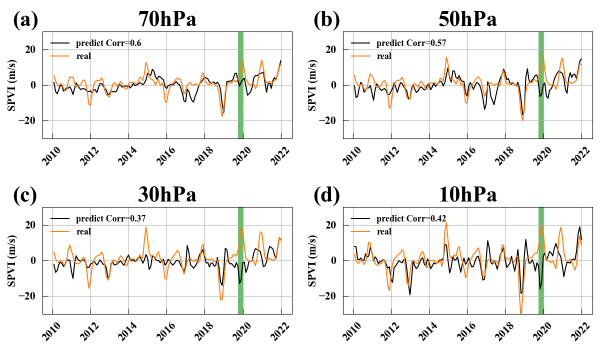}
  \caption{The predictive capability of EM4 for the Antarctic polar vortex: Panels (\textbf{a}--\textbf{d}) represent the 70 hPa, 50 hPa, 30 hPa, and 10 hPa levels, respectively, with predictive correlations of 0.60, 0.57, 0.37, and 0.42. The green vertical line indicates the time of occurrence of a sudden stratospheric warming event in September 2019.}
  \label{EM4_pred_SPVI}
\end{figure}

To quantitatively evaluate the predictive capability of EM4 for the Antarctic polar vortex, we designed a linear prediction model (see Section \ref{sec:Linear_Prediction_Model}). The prediction results are shown in Figure \ref{EM4_pred_SPVI}. The model performs best at 70 hPa and 50 hPa, achieving correlations exceeding 0.57. At 30 hPa and 10 hPa, correlations still reach above 0.37. These results indicate that the EM4 indeed has strong predictive capability for variations in the intensity of the Antarctic polar vortex. The underlying mechanisms are worthy of further analysis.

However, a significant false alarm occurred in 2019, which can be attributed to the rare sudden stratospheric warming (SSW) event in September 2019. This event is marked by a green vertical line in Figure \ref{EM4_pred_SPVI}.

Due to space limitations, the modes beyond the fourth are not further analyzed in \linebreak this paper.

\vspace{-6pt}

\vspace{-6pt}

\section{Conclusions}
\label{sec:Conclusions}
In this study, we applied the EMA to reanalysis data of zonal wind at 70--10 hPa to uncover large-scale climate patterns in the middle stratosphere. Our analysis focused on the first four leading eigen microstates (EM1–EM4), which collectively account for approximately 40--50\% of the variance.

The first eigen microstate (EM1) corresponds to the quasi-biennial oscillation (QBO) mode. Its spatial patterns reveal strong coherence in tropical stratospheric zonal winds. Its temporal evolution exhibits a periodic reversal approximately every 2.4 years. EM1’s spatial patterns accurately capture the spatial extent and signal strength of the QBO, while its temporal evolutions exhibit a strong correlation with the QBO indices.

The second eigen microstate (EM2) describes the variations in the Arctic polar vortex. Its spatial patterns are primarily concentrated in the mid-to-high latitudes of the Northern Hemisphere, with strong westerlies around 60° N–90° N and easterlies between 30° N–50° N. Its temporal evolution aligns closely with the deseasonalized Arctic polar vortex index, containing both high-frequency signals and low-frequency trends, including cycles of approximately 1.5 years, 2.4–2.6 years, and around 5--6 years. The 1.5-year cycle is linked to the influence of ENSO, while the 2.4–2.6-year cycle corresponds to the impact of QBO. The 5--6-year cycle in the Arctic polar vortex, which is rarely mentioned in existing research, potentially links to solar and volcanic activities. The EMA can automatically extract modes describing Arctic polar vortex variations from year-round data, making it easier to conduct spectral analysis and significance testing. The 5--6-year cycle of the Arctic polar vortex and its relationship with the North Pacific SST is a topic worthy of further in-depth research.

The third eigen microstate (EM3) characterizes the variations in the Antarctic polar vortex. Its spatial patterns highlight strong westerlies around 60°S. Its temporal evolution closely matches the deseasonalized Antarctic polar vortex index. Unlike EM2 (Arctic polar vortex mode), EM3 lacks a significant 5--6-year cycle, instead displaying an approximately 11-year solar cycle. During winter in each hemisphere, the polar vortex is isolated from solar radiation, with the sun's influence becoming direct only after spring begins. The Antarctic polar vortex breaks up later than the Arctic one, making it receive more solar radiation during its peak variability in late spring and early summer in the Southern Hemisphere. This leads to a more pronounced 11-year cycle.

The fourth eigen microstate (EM4) has dipole-like spatial patterns concentrated in the mid-to-high latitudes of the Southern Hemisphere. In the lower stratosphere, its temporal evolution is highly correlated with the Ni\~{n}o 3.4 index, while in the middle stratosphere, EM4's spatial patterns display a wave-1-like pattern. We found that EM4 exerts a leading influence on EM3 by about 2 months, suggesting that EM4 might be one of the factors affecting Antarctic polar vortex variations and could potentially be used to predict its intensity. We designed a linear prediction model that successfully demonstrated its predictive capability.  

The EMA is a powerful tool for studying the spatiotemporal evolution of complex systems and has the potential for application to other complex systems as well.

\vspace{6pt} 
\supplementary{\hl{The} 
 following supporting information can be downloaded at:  \linksupplementary{s1}.}

\authorcontributions{X.C., F.X., and N.T. designed the research, conceived the study, carried out the analysis, and prepared the manuscript. N.T. performed the numerical calculations and led the manuscript writing efforts. X.C., F.X., Y.Z., Y.X., X.M., H.H., H.W. generated the research ideas, discussed the results, and assisted in manuscript writing. \hl{All authors} 
 have read and agreed to the published version of the manuscript.}

\funding{\hl{This} 
 work is supported by the National Key R\&D Program of China (grant no. 2023YFE0109000) and the National Natural Science Foundation of China (grant nos. 12135003 and 12305044). This work is also supported by the ClimTip project (ClimTip contribution \#35), which has received funding from the European Union's Horizon Europe research and innovation program under grant agreement no. 101137601. As an Associated Partner, BNU has received funding from the Chinese Ministry for Science and Technology (MOST). The contents of this publication are the sole responsibility of the authors and do not necessarily reflect the opinions of the European Union or MOST.}

\dataavailability{The data that support the findings of this study are freely available. The ERA5 reanalysis dataset is available \hl{at} 
 \url{https://cds.climate.copernicus.eu/datasets/reanalysis-era5-pressure-levels-monthly-means?tab=overview} \hl{(accessed on 21 March 2025).} 
 The Ni\~{n}o 3.4 index provided by NOAA (National Oceanic and Atmospheric Administration) is available at \url{https://psl.noaa.gov/data/correlation/nina34.data} \hl{(accessed on 21 March 2025).}
} 

\acknowledgments{\hl{The} 
 authors wish to thank Niklas Boers and Deliang Chen for the valuable discussions and insightful ideas provided. The authors would also like to thank the reviewers for their valuable comments and suggestions, which have significantly contributed to improving the quality of this manuscript.}

\conflictsofinterest{The authors declare no conflicts of interest.}

\appendixtitles{yes} 
\appendixstart
\appendix
\section[\appendixname~\thesection]{\hl{Numerical Example of EMA}} \label{appA}
In this section, we present a simple numerical example of the eigen microstate approach (EMA), as outlined in \citep{sun2021eigen}. Specifically, we apply the EMA to the global surface temperature field, utilizing monthly 2-meter surface temperature data from ERA5, spanning from 1950 to 2022, with a spatial resolution of \(1^\circ \times 1^\circ\) in latitude and longitude \citep{era5citation}.
For details on the EMA, please refer to Section \ref{sec:Eigen_Microstate_Approach}. Below, we only provide a definition of the microstates of the system composed of global surface temperature and present some intuitive results from this simple numerical example. All results in this example are \linebreak from \citep{sun2021eigen} and are not part of this study.

The temperature at grid point \(i\) at a given time, \(t\), is denoted as \(T_i(t)\). To characterize the temperature fluctuations, we first subtract the temporal average at each grid point:

\[
\delta T_i(t) = T_i(t) - \langle T_i \rangle
\]
\[
\langle T_i \rangle = \frac{1}{M} \sum_{t=1}^{M} T_i(t)
\]

Here, \(\langle T_i \rangle\) represents the temporal average of \(T_i(t)\), calculated over \(M = 876 \) time steps. \( i=1,2,\dots,N \), where \( N = 181 \times 360 = 65160 \).

The standard deviation quantifies the magnitude of temperature fluctuations at each grid point \(i\). Due to the inherent differences between land and ocean, temperature fluctuations over land are typically an order of magnitude larger than those over the ocean. To eliminate this heterogeneity, temperature fluctuations normalized by their respective standard deviations can be computed. 

$$\delta S_{i}(t) = \delta T_{i}(t)/\Delta_{i} $$
$$\Delta_{i} = \sqrt{\frac{1}{M}\Sigma_{t=1}^{M}\delta T_{i}(t)^2 }$$

Then we can introduce the microstates of the Earth's surface temperature as follows: 
\[
\delta\mathbf{S(t)}= \begin{bmatrix} \delta S_{1}(t) \\ \delta S_{2}(t) \\\dots \\  \delta S_{N}(t) \end{bmatrix}
\]

Taking all the microstates at various times, we obtain an \( N \times M\) ensemble matrix of the Earth's surface temperature \(\mathbf{A}\) with the following elements: \({A}_{i}(t)=\frac{1}{\sqrt{C_{0}}} {S}_{i}(t)\), where \(\sqrt{C_0}= \sum_{i=0}^{N-1}\sum_{t=0}^{M-1}{S}_i^2(t)\).
According to EMA (see Section \ref{sec:Eigen_Microstate_Approach}), we can obtain the spatial patterns \(\mathbf{U}\) and the temporal evolutions.

\begin{figure}[H]
  \centering
  \includegraphics[scale=0.9]{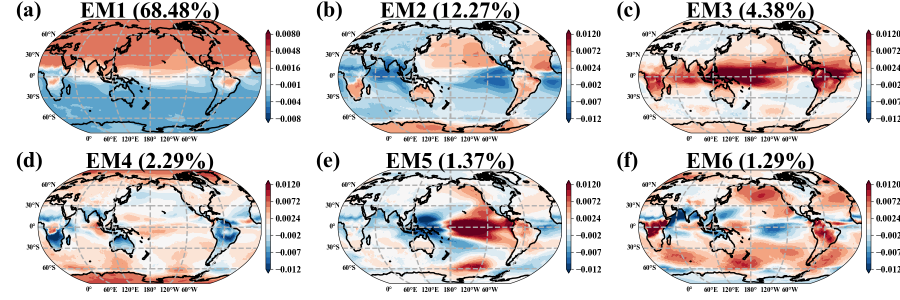}
  \caption{The spatial patterns of EM1 to EM6 (\textbf{a}–\textbf{f}). The percentages above the figure indicate \linebreak their contributions.}
  \label{EM_exmaple_U}
\end{figure}

\begin{figure}[H]
  \centering
  \includegraphics[scale=0.9]{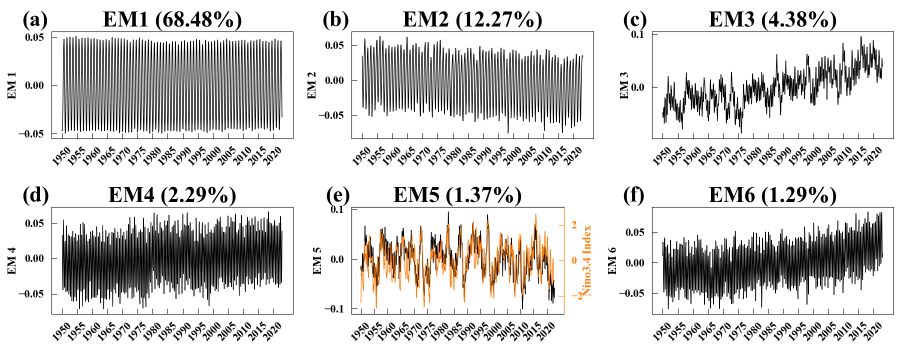}
  \caption{The temporal evolutions of EM1 to EM6 (\textbf{a}–\textbf{f}). The percentages above the figure indicate their contributions.}
  \label{EM_exmaple_V}
\end{figure}

The results are presented in Figure \ref{EM_exmaple_U} and Figure \ref{EM_exmaple_V}, corresponding to the spatial patterns and temporal evolutions, respectively.

\vspace{-6pt}

EM1, whose spatial pattern exhibits a clear inter-hemispheric contrast, represents a classic solstitial mode. EM2 reflects a significant land--sea temperature contrast, which is considered the primary driver of monsoons. EM3 captures tropical convection processes, with its power spectral density revealing 0.5-year and 1-year temporal evolution cycles. EM4 is identified as a semiannual oscillation mode, exhibiting strong semiannual signals in both tropical and mid-high latitudes. EM5 is associated with the ENSO phenomenon, although it accounts for only 1.37\% of the total variance. The temporal evolution of EM5 is compared with the Ni\~{n}o 3.4 index (orange line in Figure \ref{EM_exmaple_V}e), showing a significant correlation (\(R = 0.7\)) after 3-month averaging. 

All the mode analyses in Figure \ref{EM_exmaple_U} and Figure \ref{EM_exmaple_V} are from \citep{sun2021eigen}, where Sun et al. utilized daily surface air temperature (SAT) (2 m) data from the NCEP-NCAR reanalysis. For simplicity in this numerical example, we used monthly data. Despite minor differences in the datasets, the results remain comparable.

\begin{figure}[H]
  \centering
  \includegraphics[scale=0.9]{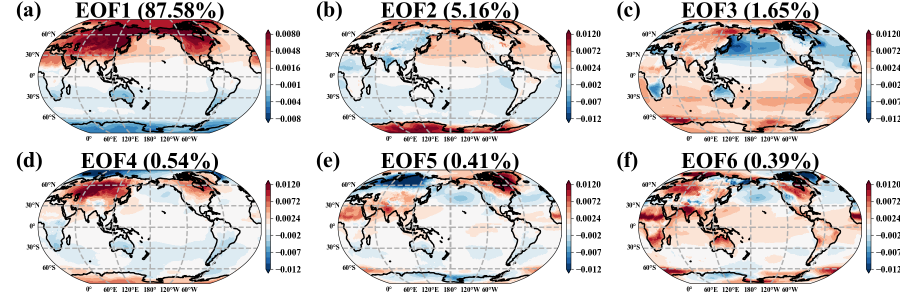}
  \caption{The spatial patterns of EOF1 to EOF6 (\textbf{a}–\textbf{f}). The percentages above the figure indicate their contributions.}
  \label{EOF_exmaple_U}
\end{figure}

To underscore the importance of properly defining microstates, we compare the results shown in Figure \ref{EM_exmaple_U} with those obtained without dividing by the standard deviation \linebreak (Figure \ref{EOF_exmaple_U}). The latter case corresponds to the commonly used empirical orthogonal function (EOF) method. As shown in Figure \ref{EOF_exmaple_U}, the spatial patterns are dominated by land, conveying less essential information.

The global temperature field is a classic example where the EMA can be effectively applied within the Earth system. Beyond this application, the EMA is well-suited for analyzing complex systems with multiple variables. For instance, when studying phase transitions of the K\'{a}rm\'{a}n vortex in turbulent systems, defining microstates that incorporate both velocity and density information provides a better representation of the system's essential characteristics \citep{li2024exploring}.

\begin{adjustwidth}{-\extralength}{0cm}

\reftitle{\highlighting{References} 
}

\PublishersNote{}
\end{adjustwidth}

\begin{thebibliography}{999}

\bibitem[Baldwin and Dunkerton(2001)]{baldwin2001stratospheric}
Baldwin, M.P.; Dunkerton, T.J.
\newblock Stratospheric harbingers of anomalous weather regimes.
\newblock {\em Science} {\bf 2001}, {\em 294},~581--584.

\bibitem[Domeisen et~al.(2020)Domeisen, Butler, Charlton-Perez,
  Ayarzag{\"u}ena, Baldwin, Dunn-Sigouin, Furtado, Garfinkel, Hitchcock,
  Karpechko, et~al.]{domeisen2020role}
Domeisen, D.I.; Butler, A.H.; Charlton-Perez, A.J.; Ayarzag{\"u}ena, B.;
  Baldwin, M.P.; Dunn-Sigouin, E.; Furtado, J.C.; Garfinkel, C.I.; Hitchcock,
  P.; Karpechko, A.Y.;  et~al.
\newblock The role of the stratosphere in subseasonal to seasonal prediction:
  2. Predictability arising from stratosphere-troposphere coupling.
\newblock {\em J. Geophys. Res. Atmos.} {\bf 2020}, {\em
  125},~e2019JD030923.

\bibitem[Baldwin et~al.(2019)Baldwin, Birner, Brasseur, Burrows, Butchart,
  Garcia, Geller, Gray, Hamilton, Harnik, et~al.]{baldwin2019100}
Baldwin, M.P.; Birner, T.; Brasseur, G.; Burrows, J.; Butchart, N.; Garcia, R.;
  Geller, M.; Gray, L.; Hamilton, K.; Harnik, N.;  et~al.
\newblock 100 years of progress in understanding the stratosphere and
  mesosphere.
\newblock {\em Meteorol. Monogr.} {\bf 2019}, {\em 59},~\hl{27.1--27.62.} 

\bibitem[Brewer(1949)]{brewer1949evidence}
Brewer, A.
\newblock Evidence for a world circulation provided by the measurements of
  helium and water vapour distribution in the stratosphere.
\newblock {\em Q. J. R. Meteorol. Soc.} {\bf
  1949}, {\em 75},~351--363.

\bibitem[Dobson(1956)]{dobson1956origin}
Dobson, G.M.B.
\newblock Origin and distribution of the polyatomic molecules in the
  atmosphere.
\newblock {\em Proc. R. Soc. Lond. Ser. A   Math. Phys. Sci.} {\bf 1956}, {\em 236},~187--193.

\bibitem[Butchart(2014)]{butchart2014brewer}
Butchart, N.
\newblock The Brewer-Dobson circulation.
\newblock {\em Rev. Geophys.} {\bf 2014}, {\em 52},~157--184.

\bibitem[Scherhag()]{scherhag1952explosionsartige}
Scherhag, R.
\newblock {Die explosionsartige Stratosphärenerwärmung des Späitwinters 1951/52.} In 
\newblock {\em Berichte des Deutschen Wetterdienstes in der US-Zone}; \hl{Deutscher Wetterdienst in der US-Zone: Bad Kissingen, Germany, } 
{1952}, \hl{Volume} 38, pp. 51--63.




\bibitem[Matsuno(1971)]{matsuno1971dynamical}
Matsuno, T.
\newblock A dynamical model of the stratospheric sudden warming.
\newblock {\em J. Atmos. Sci.} {\bf 1971}, {\em
  28},~1479--1494.

\bibitem[Labitzke(1981)]{labitzke1981stratospheric}
Labitzke, K.
\newblock Stratospheric-mesospheric midwinter disturbances: A summary of
  observed characteristics.
\newblock {\em J. Geophys. Res. Oceans} {\bf 1981}, {\em
  86},~9665--9678.

\bibitem[Baldwin et~al.(2021)Baldwin, Ayarzag{\"u}ena, Birner, Butchart,
  Butler, Charlton-Perez, Domeisen, Garfinkel, Garny, Gerber,
  et~al.]{baldwin2021sudden}
Baldwin, M.P.; Ayarzag{\"u}ena, B.; Birner, T.; Butchart, N.; Butler, A.H.;
  Charlton-Perez, A.J.; Domeisen, D.I.; Garfinkel, C.I.; Garny, H.; Gerber,
  E.P.;  et~al.
\newblock Sudden stratospheric warmings.
\newblock {\em Rev. Geophys.} {\bf 2021}, {\em 59},~e2020RG000708.

\bibitem[Ebdon(1960)]{ebdon1960notes}
Ebdon, R.
\newblock Notes on the wind flow at 50 mb in tropical and sub-tropical regions
  in January 1957 and January 1958.
\newblock {\em Q. J. R. Meteorol. Soc.} {\bf
  1960}, {\em 86},~540--542.

\bibitem[Reed et~al.(1961)Reed, Campbell, Rasmussen, and
  Rogers]{reed1961evidence}
Reed, R.J.; Campbell, W.J.; Rasmussen, L.A.; Rogers, D.G.
\newblock Evidence of a downward-propagating, annual wind reversal in the
  equatorial stratosphere.
\newblock {\em J. Geophys. Res.} {\bf 1961}, {\em 66},~813--818.

\bibitem[Lindzen and Holton(1968)]{lindzen1968theory}
Lindzen, R.S.; Holton, J.R.
\newblock A theory of the quasi-biennial oscillation.
\newblock {\em J. Atmos. Sci.} {\bf 1968}, {\em
  25},~1095--1107.

\bibitem[Holton and Lindzen(1972)]{holton1972updated}
Holton, J.R.; Lindzen, R.S.
\newblock An updated theory for the quasi-biennial cycle of the tropical
  stratosphere.
\newblock {\em J. Atmos. Sci.} {\bf 1972}, {\em
  29},~1076--1080.

\bibitem[Baldwin et~al.(2001)Baldwin, Gray, Dunkerton, Hamilton, Haynes,
  Randel, Holton, Alexander, Hirota, Horinouchi, et~al.]{baldwin2001quasi}
Baldwin, M.; Gray, L.; Dunkerton, T.; Hamilton, K.; Haynes, P.; Randel, W.;
  Holton, J.; Alexander, M.; Hirota, I.; Horinouchi, T.;  et~al.
\newblock The quasi-biennial oscillation.
\newblock {\em Rev. Geophys.} {\bf 2001}, {\em 39},~179--229.

\bibitem[Eliassen(1961)]{eliassen1961transfer}
Eliassen, A.
\newblock On the transfer of energy in stationary mountain waves.
\newblock {\em Geofys. Publ.} {\bf 1961}, {\em 22},~1--23.

\bibitem[Andrews and McIntyre(1978)]{andrews1978exact}
Andrews, D.; McIntyre, M.
\newblock An exact theory of nonlinear waves on a Lagrangian-mean flow.
\newblock {\em J. Fluid Mech.} {\bf 1978}, {\em 89},~609--646.

\bibitem[Boyd(1976)]{boyd1976noninteraction}
Boyd, J.P.
\newblock The noninteraction of waves with the zonally averaged flow on a
  spherical earth and the interrelationships on eddy fluxes of energy, heat and
  momentum.
\newblock {\em J. Atmos. Sci.} {\bf 1976}, {\em
  33},~2285--2291.

\bibitem[Marks and Eckermann(1995)]{marks1995three}
Marks, C.J.; Eckermann, S.D.
\newblock A three-dimensional nonhydrostatic ray-tracing model for gravity
  waves: Formulation and preliminary results for the middle atmosphere.
\newblock {\em J. Atmos. Sci.} {\bf 1995}, {\em
  52},~1959--1984.

\bibitem[Alexander(1997)]{alexander1997model}
Alexander, M.J.
\newblock A model of non-stationary gravity waves in the stratosphere and
  comparison to observations. In {\em Gravity Wave Processes: Their
  Parameterization in Global Climate Models}; Springer: \hl{Berlin/Heidelberg, Germany, } 
 1997; pp. 153--168.

\bibitem[Alexander and Dunkerton(1999)]{alexander1999spectral}
Alexander, M.; Dunkerton, T.
\newblock A spectral parameterization of mean-flow forcing due to breaking
  gravity waves.
\newblock {\em J. Atmos. Sci.} {\bf 1999}, {\em
  56},~4167--4182.

\bibitem[Geller et~al.(2013)Geller, Alexander, Love, Bacmeister, Ern, Hertzog,
  Manzini, Preusse, Sato, Scaife, et~al.]{geller2013comparison}
Geller, M.A.; Alexander, M.J.; Love, P.T.; Bacmeister, J.; Ern, M.; Hertzog,
  A.; Manzini, E.; Preusse, P.; Sato, K.; Scaife, A.A.;  et~al.
\newblock A comparison between gravity wave momentum fluxes in observations and
  climate models.
\newblock {\em J. Clim.} {\bf 2013}, {\em 26},~6383--6405.

\bibitem[Baldwin et~al.(2024)Baldwin, Birner, and
  Ayarzag{\"u}ena]{baldwin2024tropospheric}
Baldwin, M.P.; Birner, T.; Ayarzag{\"u}ena, B.
\newblock Tropospheric amplification of stratosphere--troposphere coupling.
\newblock {\em Q. J. R. Meteorol. Soc.} {\bf
  2024}, \hl{\emph{150}, 5188--5205.}

\bibitem[Hall and Plumb(1994)]{hall1994age}
Hall, T.M.; Plumb, R.A.
\newblock Age as a diagnostic of stratospheric transport.
\newblock {\em J. Geophys. Res. Atmos.} {\bf 1994}, {\em
  99},~1059--1070.

\bibitem[Dobson(1931)]{dobson1931photoelectric}
Dobson, G.
\newblock A photoelectric spectrophotometer for measuring the amount of
  atmospheric ozone.
\newblock {\em Proc. Phys. Soc.} {\bf 1931}, {\em 43},~324.

\bibitem[Bates and Nicolet(1950)]{bates1950photochemistry}
Bates, D.R.; Nicolet, M.
\newblock The photochemistry of atmospheric water vapor.
\newblock {\em J. Geophys. Res.} {\bf 1950}, {\em 55},~301--327.

\bibitem[Crutzen(1970)]{crutzen1970influence}
Crutzen, P.J.
\newblock The Influence of Nitrogen Oxides on Atmospheric Ozone Content. In {\em Paul J. Crutzen: A Pioneer on Atmospheric Chemistry and Climate Change in the Anthropocene}; Crutzen, P.J., Brauch, H.G., Eds.; Springer: Berlin/Heidelberger, Germany, 2016; \hl{Volume 50.} 



\emph{Q.  J.
  R. Meteorol. Soc.} \textbf{1970}, \emph{96}, 320--325.

\bibitem[Stolarski and Cicerone(1974)]{stolarski1974stratospheric}
Stolarski, R.S.; Cicerone, R.J.
\newblock Stratospheric chlorine: A possible sink for ozone.
\newblock {\em Can. J. Chem.} {\bf 1974}, {\em 52},~1610--1615.

\bibitem[Wofsy et~al.(1975)Wofsy, McElroy, and Yung]{wofsy1975chemistry}
Wofsy, S.C.; McElroy, M.B.; Yung, Y.L.
\newblock The chemistry of atmospheric bromine.
\newblock {\em Geophys. Res. Lett.} {\bf 1975}, {\em 2},~215--218.

\bibitem[Johnston(1971)]{johnston1971reduction}
Johnston, H.
\newblock Reduction of stratospheric ozone by nitrogen oxide catalysts from
  supersonic transport exhaust.
\newblock {\em Science} {\bf 1971}, {\em 173},~517--522.

\bibitem[Molina and Rowland(1974)]{molina1974stratospheric}
Molina, M.J.; Rowland, F.S.
\newblock Stratospheric sink for chlorofluoromethanes: Chlorine atom-catalysed
  destruction of ozone.
\newblock {\em Nature} {\bf 1974}, {\em 249},~810--812.

\bibitem[Crutzen and Arnold(1986)]{crutzen1986nitric}
Crutzen, P.J.; Arnold, F.
\newblock Nitric acid cloud formation in the cold Antarctic stratosphere: A
  major cause for the springtime ‘ozone hole’.
\newblock {\em Nature} {\bf 1986}, {\em 324},~651--655.

\bibitem[Toon et~al.(1986)Toon, Hamill, Turco, and Pinto]{toon1986condensation}
Toon, O.B.; Hamill, P.; Turco, R.P.; Pinto, J.
\newblock Condensation of HNO3 and HCl in the winter polar stratospheres.
\newblock {\em Geophys. Res. Lett.} {\bf 1986}, {\em 13},~1284--1287.

\bibitem[Solomon et~al.(1986)Solomon, Garcia, Rowland, and
  Wuebbles]{solomon1986depletion}
Solomon, S.; Garcia, R.R.; Rowland, F.S.; Wuebbles, D.J.
\newblock On the depletion of Antarctic ozone.
\newblock {\em Nature} {\bf 1986}, {\em 321},~755--758.

\bibitem[McElroy et~al.(1986)McElroy, Salawitch, Wofsy, and
  Logan]{mcelroy1986reductions}
McElroy, M.B.; Salawitch, R.J.; Wofsy, S.C.; Logan, J.A.
\newblock Reductions of Antarctic ozone due to synergistic interactions of
  chlorine and bromine.
\newblock {\em Nature} {\bf 1986}, {\em 321},~759--762.

\bibitem[Tung et~al.(1986)Tung, Ko, Rodriguez, and Dak~Sze]{tung1986antarctic}
Tung, K.K.; Ko, M.K.; Rodriguez, J.M.; Dak~Sze, N.
\newblock Are Antarctic ozone variations a manifestation of dynamics or
  chemistry?
\newblock {\em Nature} {\bf 1986}, {\em 322},~811--814.

\bibitem[Molina and Molina(1987)]{molina1987production}
Molina, L.T.; Molina, M.J.
\newblock Production of chlorine oxide (Cl$_2$O$_2$) from the self-reaction of the
  chlorine oxide (ClO) radical.
\newblock {\em J. Phys. Chem.} {\bf 1987}, {\em 91},~433--436.

\bibitem[Ruffini and Wheeler(1971)]{ruffini1971introducing}
Ruffini, R.; Wheeler, J.A.
\newblock Introducing the black hole.
\newblock {\em Phys. Today} {\bf 1971}, {\em 24},~30--41.

\bibitem[Castelvecchi(2017)]{castelvecchi2017gravitational}
Castelvecchi, D.
\newblock Gravitational wave detection wins physics Nobel.
\newblock {\em Nature} {\bf 2017}, {\em 550}, 19. 
\newblock {\url{https://doi.org/10.1038/nature.2017.22737}}.


\bibitem[Wright(2013)]{higgsphysics}
Wright, A.
\newblock Nobel Prize 2013: Englert and Higgs.
\newblock {\em Nature Physics} {\bf 2013}, {\em 9},~692.
\newblock {\url{https://doi.org/10.1038/nphys2800}}.


\bibitem[Fan et~al.(2021)Fan, Meng, Ludescher, Chen, Ashkenazy, Kurths, Havlin,
  and Schellnhuber]{fan2021statistical}
Fan, J.; Meng, J.; Ludescher, J.; Chen, X.; Ashkenazy, Y.; Kurths, J.; Havlin,
  S.; Schellnhuber, H.J.
\newblock Statistical physics approaches to the complex Earth system.
\newblock {\em Phys. Rep.} {\bf 2021}, {\em 896},~1--84.

\bibitem[Pfeffer(1981)]{pfeffer1981wave}
Pfeffer, R.L.
\newblock Wave-mean flow interactions in the atmosphere.
\newblock {\em J. Atmos. Sci.} {\bf 1981}, {\em
  38},~1340--1359.

\bibitem[Anderson(1972)]{anderson1972more}
Anderson, P.W.
\newblock More Is Different: Broken symmetry and the nature of the hierarchical
  structure of science.
\newblock {\em Science} {\bf 1972}, {\em 177},~393--396.

\bibitem[Sun et~al.(2021)Sun, Hu, Zhang, Lu, Lu, Fan, Li, Deng, and
  Chen]{sun2021eigen}
Sun, Y.; Hu, G.; Zhang, Y.; Lu, B.; Lu, Z.; Fan, J.; Li, X.; Deng, Q.; Chen, X.
\newblock Eigen microstates and their evolutions in complex systems.
\newblock {\em Commun. Theor. Phys.} {\bf 2021}, {\em
  73},~065603.

\bibitem[Hu et~al.(2019)Hu, Liu, Liu, Chen, and Chen]{hu2019condensation}
Hu, G.; Liu, T.; Liu, M.; Chen, W.; Chen, X.
\newblock Condensation of eigen microstate in statistical ensemble and phase
  transition.
\newblock {\em Sci. China Phys. Mech. Astron.} {\bf 2019}, {\em
  62},~\hl{990511.} 

\bibitem[Li et~al.(2021)Li, Xue, Sun, Fan, Li, Liu, Han, Di, and
  Chen]{li2021discontinuous}
Li, X.; Xue, T.; Sun, Y.; Fan, J.; Li, H.; Liu, M.; Han, Z.; Di, Z.; Chen, X.
\newblock Discontinuous and continuous transitions of collective behaviors in
  living systems.
\newblock {\em Chin. Phys. B} {\bf 2021}, {\em 30},~128703.

\bibitem[Zhang et~al.(2024)Zhang, Liu, Hu, Liu, and Chen]{zhang2024eigen}
Zhang, Y.; Liu, M.; Hu, G.; Liu, T.; Chen, X.
\newblock Eigen microstates in self-organized criticality.
\newblock {\em Phys. Rev. E} {\bf 2024}, {\em 109},~044130.

\bibitem[Hu et~al.(2023)Hu, Liu, and Chen]{hu2023quantum}
Hu, G.; Liu, M.; Chen, X.
\newblock Quantum phase transition and eigen microstate condensation in the
  quantum Rabi model.
\newblock {\em Phys. A Stat. Mech. Its Appl.} {\bf
  2023}, {\em 630},~129210.

\bibitem[Li et~al.(2024)Li, Xiang, Xue, Wang, and Chen]{li2024exploring}
Li, X.; Xiang, X.; Xue, T.; Wang, L.; Chen, X.
\newblock Exploring multiple phases and first-order phase transitions in
  K{\'a}rm{\'a}n Vortex Street.
\newblock {\em Sci. China Phys. Mech. Astron.} {\bf 2024}, {\em
  67},~110511.

\bibitem[Ma et~al.(2024)Ma, Liang, Chen, Xie, Zuo, Sun, and
  Ding]{ma2024increased}
Ma, X.; Liang, R.; Chen, X.; Xie, F.; Zuo, J.; Sun, C.; Ding, R.
\newblock Increased predictability of extreme El Ni{\~n}o from decadal
  interbasin interaction. \emph{{Geophys. Res. Lett.}} \textbf{2024}, \emph{51}, e2024GL110943.


\bibitem[Wang et~al.(2024)Wang, Fan, Chen, Xie, and Wang]{wang2024holistic}
Wang, X.; Fan, H.; Chen, X.; Xie, Y.; Wang, H.
\newblock Holistic evolution of ecosystem in Heihe River Basin from the
  perspective of eigen microstates.
\newblock {\em Ecol. Indic.} {\bf 2024}, {\em 159},~111689.

\bibitem[Chen et~al.(2021)Chen, Ying, Chen, Zhang, Lu, Fan, and
  Chen]{chen2021eigen}
Chen, X.; Ying, N.; Chen, D.; Zhang, Y.; Lu, B.; Fan, J.; Chen, X.
\newblock Eigen microstates and their evolution of global ozone at different
  geopotential heights.
\newblock {\em Chaos} {\bf 2021}, {\em 31}, {071102}. {\url{https://doi.org/10.1063/5.0058599}}.

\bibitem[Chen et~al.(2023)Chen, Ren, Tang, Zhou, Zhou, Zuo, Cui, Chen, Liu, He,
  et~al.]{chen2023leading}
Chen, X.; Ren, H.; Tang, Z.; Zhou, K.; Zhou, L.; Zuo, Z.; Cui, X.; Chen, X.;
  Liu, Z.; He, Y.;  et~al.
\newblock Leading basic modes of spontaneous activity drive individual
  functional connectivity organization in the resting human brain.
\newblock {\em Commun. Biol.} {\bf 2023}, {\em 6},~892.


\bibitem[Hersbach et~al.(2023)Hersbach, Bell, Berrisford, Biavati, Horányi,
  Muñoz~Sabater, Nicolas, Peubey, Radu, Rozum, Schepers, Simmons, Soci, Dee,
  and Thépaut]{era5citation}
Hersbach, H.; Bell, B.; Berrisford, P.; Biavati, G.; Horányi, A.;
  Muñoz~Sabater, J.; Nicolas, J.; Peubey, C.; Radu, R.; Rozum, I.;  et~al.
\newblock ERA5 monthly averaged data on pressure levels from 1940 to present,
  2023.
\newblock Copernicus Climate Change Service (C3S) Climate Data Store (CDS).
Available online: \url{https://doi.org/10.24381/cds.6860a573} \hl{(accessed on 3 May 2023).} 


\bibitem[Liang et~al.(2023)Liang, Rao, Guo, Lu, and Shi]{liang2023northern}
Liang, Z.; Rao, J.; Guo, D.; Lu, Q.; Shi, C.
\newblock Northern winter stratospheric polar vortex regimes and their possible
  influence on the extratropical troposphere.
\newblock {\em Clim. Dyn.} {\bf 2023}, {\em 60},~3167--3186.

\bibitem[Yamazaki et~al.(2020)Yamazaki, Nakamura, Ukita, and
  Hoshi]{yamazaki2020tropospheric}
Yamazaki, K.; Nakamura, T.; Ukita, J.; Hoshi, K.
\newblock A tropospheric pathway of the stratospheric quasi-biennial
  oscillation (QBO) impact on the boreal winter polar vortex.
\newblock {\em Atmos. Chem. Phys.} {\bf 2020}, {\em
  20},~5111--5127.

\bibitem[Luo et~al.(2023)Luo, Luo, Xie, Tian, Hu, Yuan, Zhang, and
  Wang]{luo2023key}
Luo, F.; Luo, J.; Xie, F.; Tian, W.; Hu, Y.; Yuan, L.; Zhang, R.; Wang, T.
\newblock The Key Role of the Vertical Structure of the Stratospheric
  Quasi-Biennial Oscillation in the Variations of Asian Precipitation in
  Summer.
\newblock {\em Geophys. Res. Lett.} {\bf 2023}, {\em
  50},~e2023GL105863.

\bibitem[Rayner et~al.(2003)Rayner, Parker, Horton, Folland, Alexander, Rowell,
  Kent, and Kaplan]{Rayner2003}
Rayner, N.A.; Parker, D.E.; Horton, E.B.; Folland, C.K.; Alexander, L.V.;
  Rowell, D.P.; Kent, E.C.; Kaplan, A.
\newblock Global analyses of sea surface temperature, sea ice, and night marine
  air temperature since the late nineteenth century.
\newblock {\em J. Geophys. Res.} {\bf 2003}, {\em 108},~4407.
\newblock {\url{https://doi.org/10.1029/2002JD002670}}.

\bibitem[Ross(2017)]{ross2017introductory}
Ross, S.M.
\newblock {\em Introductory Statistics}; Academic Press: \hl{Cambridge, MA, USA},  2017.

\bibitem[Guez et~al.(2014)Guez, Gozolchiani, and Havlin]{guez2014influence}
Guez, O.C.; Gozolchiani, A.; Havlin, S.
\newblock Influence of autocorrelation on the topology of the climate network.
\newblock {\em Phys. Rev. E} {\bf 2014}, {\em 90},~062814.


\bibitem[Lancaster et~al.(2018)Lancaster, Iatsenko, Pidde, Ticcinelli, and
  Stefanovska]{lancaster2018surrogate}
Lancaster, G.; Iatsenko, D.; Pidde, A.; Ticcinelli, V.; Stefanovska, A.
\newblock Surrogate data for hypothesis testing of physical systems.
\newblock {\em Physics Reports} {\bf 2018}, {\em 748},~1--60.
\bibitem[Theiler et~al.(1992)Theiler, Eubank, Longtin, Galdrikian, and
  Farmer]{theiler1992testing}
Theiler, J.; Eubank, S.; Longtin, A.; Galdrikian, B.; Farmer, J.D.
\newblock Testing for nonlinearity in time series: the method of surrogate
  data.
\newblock {\em Physica D: Nonlinear Phenomena} {\bf 1992}, {\em 58},~77--94.
\bibitem[Gilman et~al.(1963)Gilman, Fuglister, and Mitchell]{Gilman1963}
Gilman, D.L.; Fuglister, F.J.; Mitchell, J.M.
\newblock On the Power Spectrum of "Red Noise".
\newblock {\em Journal of Atmospheric Sciences} {\bf 1963}, {\em 20},~182--184.
\newblock
  {\url{https://doi.org/10.1175/1520-0469(1963)020<0182:OTPSON>2.0.CO;2}}.



\bibitem[Mohanakumar(2008)]{mohanakumar2008stratosphere}
Mohanakumar, K.
\newblock {\em Stratosphere Troposphere Interactions: An Introduction};
  Springer Science \& Business Media: \hl{Berlin/Heidelberg, Germany,}  2008.

\bibitem[Kawatani et~al.(2010)Kawatani, Watanabe, Sato, Dunkerton, Miyahara,
  and Takahashi]{kawatani2010roles}
Kawatani, Y.; Watanabe, S.; Sato, K.; Dunkerton, T.J.; Miyahara, S.; Takahashi,
  M.
\newblock The roles of equatorial trapped waves and internal inertia--gravity
  waves in driving the quasi-biennial oscillation. Part I: Zonal mean wave
  forcing.
\newblock {\em J. Atmos. Sci.} {\bf 2010}, {\em
  67},~963--980.

\bibitem[Richter et~al.(2014)Richter, Solomon, and
  Bacmeister]{richter2014simulation}
Richter, J.H.; Solomon, A.; Bacmeister, J.T.
\newblock On the simulation of the quasi-biennial oscillation in the Community
  Atmosphere Model, version 5.
\newblock {\em J. Geophys. Res. Atmos.} {\bf 2014}, {\em
  119},~3045--3062.

\bibitem[Kim and Chun(2015{\natexlab{a}})]{kim2015momentum}
Kim, Y.H.; Chun, H.Y.
\newblock Momentum forcing of the quasi-biennial oscillation by equatorial
  waves in recent reanalyses.
\newblock {\em Atmos. Chem. Phys.} {\bf 2015}, {\em
  15},~6577--6587.

\bibitem[Kim and Chun(2015{\natexlab{b}})]{kim2015contributions}
Kim, Y.H.; Chun, H.Y.
\newblock Contributions of equatorial wave modes and parameterized gravity
  waves to the tropical QBO in HadGEM2.
\newblock {\em J. Geophys. Res. Atmos.} {\bf 2015}, {\em
  120},~1065--1090.

\bibitem[Pahlavan et~al.(2021)Pahlavan, Fu, Wallace, and
  Kiladis]{pahlavan2021revisiting}
Pahlavan, H.A.; Fu, Q.; Wallace, J.M.; Kiladis, G.N.
\newblock Revisiting the quasi-biennial oscillation as seen in ERA5. Part I:
  Description and momentum budget.
\newblock {\em J. Atmos. Sci.} {\bf 2021}, {\em
  78},~673--691.

\bibitem[Ern et~al.(2014)Ern, Ploeger, Preusse, Gille, Gray, Kalisch, Mlynczak,
  Russell~III, and Riese]{ern2014interaction}
Ern, M.; Ploeger, F.; Preusse, P.; Gille, J.; Gray, L.; Kalisch, S.; Mlynczak,
  M.; Russell~III, J.; Riese, M.
\newblock Interaction of gravity waves with the QBO: A satellite perspective.
\newblock {\em J. Geophys. Res. Atmos.} {\bf 2014}, {\em
  119},~2329--2355.

\bibitem[Newman et~al.(2016)Newman, Coy, Pawson, and Lait]{newman2016anomalous}
Newman, P.; Coy, L.; Pawson, S.; Lait, L.
\newblock The anomalous change in the QBO in 2015--2016.
\newblock {\em Geophys. Res. Lett.} {\bf 2016}, {\em 43},~8791--8797.

\bibitem[Saunders et~al.(2020)Saunders, Lea, and Smallwood]{saunders2020quasi}
Saunders, M.A.; Lea, A.S.; Smallwood, J.R.
\newblock The quasi-biennial oscillation: A second disruption in four years.
\newblock {\em ESS Open Archive} {\bf {2020}}, {\url{https://doi.org/10.1002/essoar.10504326.1}}.

\bibitem[Wang et~al.(2023)Wang, Rao, Lu, Ju, Yang, and Luo]{wang2023revisit}
Wang, Y.; Rao, J.; Lu, Y.; Ju, Z.; Yang, J.; Luo, J.
\newblock A revisit and comparison of the quasi-biennial oscillation (QBO)
  disruption events in 2015/16 and 2019/20.
\newblock {\em Atmos. Res.} {\bf 2023}, \emph{\hl{294}}, 106970.

\bibitem[Osprey et~al.(2016)Osprey, Butchart, Knight, Scaife, Hamilton, Anstey,
  Schenzinger, and Zhang]{osprey2016unexpected}
Osprey, S.M.; Butchart, N.; Knight, J.R.; Scaife, A.A.; Hamilton, K.; Anstey,
  J.A.; Schenzinger, V.; Zhang, C.
\newblock An unexpected disruption of the atmospheric quasi-biennial
  oscillation.
\newblock {\em Science} {\bf 2016}, {\em 353},~1424--1427.

\bibitem[Kang et~al.(2022)Kang, Chun, Son, Garcia, An, and Park]{kang2022role}
Kang, M.J.; Chun, H.Y.; Son, S.W.; Garcia, R.R.; An, S.I.; Park, S.H.
\newblock Role of tropical lower stratosphere winds in quasi-biennial
  oscillation disruptions.
\newblock {\em Sci. Adv.} {\bf 2022}, {\em 8},~eabm7229.

\bibitem[Anstey et~al.(2021)Anstey, Banyard, Butchart, Coy, Newman, Osprey, and
  Wright]{anstey2021prospect}
Anstey, J.A.; Banyard, T.P.; Butchart, N.; Coy, L.; Newman, P.A.; Osprey, S.;
  Wright, C.J.
\newblock Prospect of increased disruption to the QBO in a changing climate.
\newblock {\em Geophys. Res. Lett.} {\bf 2021}, {\em
  48},~e2021GL093058.

\bibitem[Holton and Tan(1980)]{holton1980influence}
Holton, J.R.; Tan, H.C.
\newblock The influence of the equatorial quasi-biennial oscillation on the
  global circulation at 50 mb.
\newblock {\em J. Atmos. Sci.} {\bf 1980}, {\em
  37},~2200--2208.

\bibitem[Anstey and Shepherd(2014)]{anstey2014high}
Anstey, J.A.; Shepherd, T.G.
\newblock High-latitude influence of the quasi-biennial oscillation.
\newblock {\em Q. J. R. Meteorol. Soc.} {\bf
  2014}, {\em 140},~1--21.

\bibitem[Garfinkel et~al.(2018)Garfinkel, Schwartz, Domeisen, Son, Butler, and
  White]{garfinkel2018extratropical}
Garfinkel, C.I.; Schwartz, C.; Domeisen, D.I.; Son, S.W.; Butler, A.H.; White,
  I.P.
\newblock Extratropical atmospheric predictability from the quasi-biennial
  oscillation in subseasonal forecast models.
\newblock {\em J. Geophys. Res. Atmos.} {\bf 2018}, {\em
  123},~7855--7866.

\bibitem[Rao et~al.(2020)Rao, Garfinkel, and White]{rao2020impact}
Rao, J.; Garfinkel, C.I.; White, I.P.
\newblock Impact of the quasi-biennial oscillation on the northern winter
  stratospheric polar vortex in CMIP5/6 models.
\newblock {\em J.  Clim.} {\bf 2020}, {\em 33},~4787--4813.

\bibitem[Randel and Wu(1996)]{randel1996isolation}
Randel, W.J.; Wu, F.
\newblock Isolation of the ozone QBO in SAGE II data by singular-value
  decomposition.
\newblock {\em Journal of Atmospheric Sciences} {\bf 1996}, {\em
  53},~2546--2559.



\bibitem[Rao et~al.(2019)Rao, Yu, Guo, Shi, Chen, and Hu]{rao2019evaluating}
Rao, J.; Yu, Y.; Guo, D.; Shi, C.; Chen, D.; Hu, D.
\newblock Evaluating the Brewer--Dobson circulation and its responses to ENSO,
  QBO, and the solar cycle in different reanalyses.
\newblock {\em Earth Planet. Phys.} {\bf 2019}, {\em 3},~166--181.

\bibitem[Garfinkel and Hartmann(2011)]{garfinkel2011influence}
Garfinkel, C.I.; Hartmann, D.L.
\newblock The influence of the quasi-biennial oscillation on the troposphere in
  winter in a hierarchy of models. Part II: Perpetual winter WACCM runs.
\newblock {\em J. Atmos. Sci.} {\bf 2011}, {\em
  68},~2026--2041.

\bibitem[Wang et~al.(2018)Wang, Kim, and Chang]{wang2018interannual}
Wang, J.; Kim, H.M.; Chang, E.K.
\newblock Interannual modulation of Northern Hemisphere winter storm tracks by
  the QBO.
\newblock {\em Geophys. Res. Lett.} {\bf 2018}, {\em 45},~2786--2794.

\bibitem[Martin et~al.(2021)Martin, Son, Butler, Hendon, Kim, Sobel, Yoden, and
  Zhang]{martin2021influence}
Martin, Z.; Son, S.W.; Butler, A.; Hendon, H.; Kim, H.; Sobel, A.; Yoden, S.;
  Zhang, C.
\newblock The influence of the quasi-biennial oscillation on the Madden--Julian
  oscillation.
\newblock {\em Nat. Rev. Earth Environ.} {\bf 2021}, {\em
  2},~477--489.

\bibitem[Anstey et~al.(2022)Anstey, Osprey, Alexander, Baldwin, Butchart, Gray,
  Kawatani, Newman, and Richter]{anstey2022impacts}
Anstey, J.A.; Osprey, S.M.; Alexander, J.; Baldwin, M.P.; Butchart, N.; Gray,
  L.; Kawatani, Y.; Newman, P.A.; Richter, J.H.
\newblock Impacts, processes and projections of the quasi-biennial oscillation.
\newblock {\em Nat. Rev. Earth Environ.} {\bf 2022}, {\em
  3},~588--603.

\bibitem[Zhang et~al.(2024)Zhang, Zhou, Tian, Zhang, Zhang, and
  Luo]{zhang2024stratospheric}
Zhang, R.; Zhou, W.; Tian, W.; Zhang, Y.; Zhang, J.; Luo, J.
\newblock A stratospheric precursor of East Asian summer droughts and floods.
\newblock {\em Nat. Commun.} {\bf 2024}, {\em 15},~247.

\bibitem[Wang et~al.(2019)Wang, Tippett, Sobel, Martin, and
  Vitart]{wang2019impact}
Wang, S.; Tippett, M.K.; Sobel, A.H.; Martin, Z.K.; Vitart, F.
\newblock Impact of the QBO on prediction and predictability of the MJO
  convection.
\newblock {\em J. Geophys. Res. Atmos.} {\bf 2019}, {\em
  124},~11766--11782.

\bibitem[Naujokat(1986)]{naujokat1986update}
Naujokat, B.
\newblock An update of the observed quasi-biennial oscillation of the
  stratospheric winds over the tropics.
\newblock {\em J. Atmos. Sci.} {\bf 1986}, {\em
  43},~1873--1877.

\bibitem[Dunkerton and Delisi(1985)]{dunkerton1985climatology}
Dunkerton, T.J.; Delisi, D.P.
\newblock Climatology of the equatorial lower stratosphere.
\newblock {\em J. Atmos. Sci.} {\bf 1985}, {\em
  42},~376--396.

\bibitem[Xu and Ren(2023)]{xu2023ceof}
Xu, W.; Ren, H.L.
\newblock A CEOF-based method for measuring amplitude and phase properties of
  the QBO.
\newblock {\em Clim. Dyn.} {\bf 2023}, {\em 61},~923--937.


\bibitem[Waugh et~al.(2017)Waugh, Sobel, and Polvani]{waugh2017polar}
Waugh, D.W.; Sobel, A.H.; Polvani, L.M.
\newblock What is the polar vortex and how does it influence weather?
\newblock {\em Bull. Am. Meteorol. Soc.} {\bf 2017},
  {\em 98},~37--44.

\bibitem[Zuev et~al.(2021)Zuev, Savelieva, Borovko, and
  Krupchatnikov]{zuev2021influence}
Zuev, V.V.; Savelieva, E.; Borovko, I.V.; Krupchatnikov, V.N.
\newblock Influence of the subtropical stratosphere on the Antarctic polar
  vortex during spring 2019.
\newblock In Proceedings of the 27th International Symposium on Atmospheric and
  Ocean Optics, Atmospheric Physics, SPIE,  \hl{Moscow, Russia, 5--9 July 2021; }Volume 11916, pp. 1539--1544.

\bibitem[Schoeberl and Hartmann(1991)]{schoeberl1991dynamics}
Schoeberl, M.R.; Hartmann, D.L.
\newblock The dynamics of the stratospheric polar vortex and its relation to
  springtime ozone depletions.
\newblock {\em Science} {\bf 1991}, {\em 251},~46--52.

\bibitem[Roy and Kuttippurath(2022)]{roy2022dynamical}
Roy, R.; Kuttippurath, J.
\newblock The dynamical evolution of Sudden Stratospheric Warmings of the
  Arctic winters in the past decade 2011--2021.
\newblock {\em SN Appl. Sci.} {\bf 2022}, {\em 4},~105.

\bibitem[Thompson et~al.(2002)Thompson, Baldwin, and
  Wallace]{thompson2002stratospheric}
Thompson, D.W.; Baldwin, M.P.; Wallace, J.M.
\newblock Stratospheric connection to Northern Hemisphere wintertime weather:
  Implications for prediction.
\newblock {\em J.  Clim.} {\bf 2002}, {\em 15},~1421--1428.

\bibitem[Lim et~al.(2019)Lim, Hendon, Boschat, Hudson, Thompson, Dowdy, and
  Arblaster]{lim2019australian}
Lim, E.P.; Hendon, H.H.; Boschat, G.; Hudson, D.; Thompson, D.W.; Dowdy, A.J.;
  Arblaster, J.M.
\newblock Australian hot and dry extremes induced by weakenings of the
  stratospheric polar vortex.
\newblock {\em Nat. Geosci.} {\bf 2019}, {\em 12},~896--901.

\bibitem[Kolstad et~al.(2010)Kolstad, Breiteig, and
  Scaife]{kolstad2010association}
Kolstad, E.W.; Breiteig, T.; Scaife, A.A.
\newblock The association between stratospheric weak polar vortex events and
  cold air outbreaks in the Northern Hemisphere.
\newblock {\em Q. J. R. Meteorol. Soc.} {\bf
  2010}, {\em 136},~886--893.

\bibitem[Overland et~al.(2020)Overland, Hall, Hanna, Karpechko, Vihma, Wang,
  and Zhang]{overland2020polar}
Overland, J.; Hall, R.; Hanna, E.; Karpechko, A.; Vihma, T.; Wang, M.; Zhang,
  X.
\newblock The polar vortex and extreme weather: The Beast from the East in
  winter 2018.
\newblock {\em Atmosphere} {\bf 2020}, {\em 11},~664.

\bibitem[Yu et~al.(2018)Yu, Cai, Shi, and Ren]{yu2018linkage}
Yu, Y.; Cai, M.; Shi, C.; Ren, R.
\newblock On the linkage among strong stratospheric mass circulation,
  stratospheric sudden warming, and cold weather events.
\newblock {\em Mon. Weather. Rev.} {\bf 2018}, {\em 146},~2717--2739.

\bibitem[Rao et~al.(2018)Rao, Ren, Chen, Yu, and Zhou]{rao2018stratospheric}
Rao, J.; Ren, R.; Chen, H.; Yu, Y.; Zhou, Y.
\newblock The stratospheric sudden warming event in February 2018 and its
  prediction by a climate system model.
\newblock {\em J. Geophys. Res. Atmos.} {\bf 2018}, {\em
  123},~13--332.

\bibitem[Bao et~al.(2017)Bao, Tan, Hartmann, and Ceppi]{bao2017classifying}
Bao, M.; Tan, X.; Hartmann, D.L.; Ceppi, P.
\newblock Classifying the tropospheric precursor patterns of sudden
  stratospheric warmings.
\newblock {\em Geophys. Res. Lett.} {\bf 2017}, {\em 44},~8011--8016.

\bibitem[Ma et~al.(2022)Ma, Yang, Tan, and Bao]{ma2022possible}
Ma, C.; Yang, P.; Tan, X.; Bao, M.
\newblock Possible causes of the occurrence of a rare Antarctic sudden
  stratospheric warming in 2019.
\newblock {\em Atmosphere} {\bf 2022}, {\em 13},~147.

\bibitem[Wei et~al.(2007)Wei, Chen, and Huang]{wei2007dynamical}
Wei, K.; Chen, W.; Huang, R.
\newblock Dynamical diagnosis of the breakup of the stratospheric polar vortex
  in the Northern Hemisphere.
\newblock {\em Sci. China Ser. D Earth Sci.} {\bf 2007}, {\em
  50},~1369--1379.

\bibitem[Lu et~al.(2014)Lu, Bracegirdle, Phillips, Bushell, and
  Gray]{lu2014mechanisms}
Lu, H.; Bracegirdle, T.J.; Phillips, T.; Bushell, A.; Gray, L.
\newblock Mechanisms for the Holton-Tan relationship and its decadal variation.
\newblock {\em J. Geophys. Res. Atmos.} {\bf 2014}, {\em
  119},~2811--2830.

\bibitem[Yang et~al.(2018)Yang, Li, Yu, Hu, Dong, and He]{yang2018Nino}
Yang, S.; Li, Z.; Yu, J.Y.; Hu, X.; Dong, W.; He, S.
\newblock El Ni{\~n}o--Southern Oscillation and its impact in the changing
  climate.
\newblock {\em Natl. Sci. Rev.} {\bf 2018}, {\em 5},~840--857.

\bibitem[Domeisen et~al.(2019)Domeisen, Garfinkel, and
  Butler]{domeisen2019teleconnection}
Domeisen, D.I.; Garfinkel, C.I.; Butler, A.H.
\newblock The teleconnection of El Ni{\~n}o Southern Oscillation to the
  stratosphere.
\newblock {\em Rev. Geophys.} {\bf 2019}, {\em 57},~5--47.

\bibitem[Jevrejeva et~al.(2004)Jevrejeva, Moore, and
  Grinsted]{jevrejeva2004oceanic}
Jevrejeva, S.; Moore, J.; Grinsted, A.
\newblock Oceanic and atmospheric transport of multiyear El Nino--Southern
  Oscillation (ENSO) signatures to the polar regions.
\newblock {\em Geophys. Res. Lett.} {\bf 2004}, {\em \hl{31}
}.  https://doi.org/10.1029/2004GL020871.

\bibitem[Hu et~al.(2018)Hu, Guan, Tian, and Ren]{hu2018recent}
Hu, D.; Guan, Z.; Tian, W.; Ren, R.
\newblock Recent strengthening of the stratospheric Arctic vortex response to
  warming in the central North Pacific.
\newblock {\em Nat. Commun.} {\bf 2018}, {\em 9},~1697.

\bibitem[Li et~al.(2018)Li, Tian, Xie, Wen, Zhang, Hu, and
  Han]{li2018connection}
Li, Y.; Tian, W.; Xie, F.; Wen, Z.; Zhang, J.; Hu, D.; Han, Y.
\newblock The connection between the second leading mode of the winter North
  Pacific sea surface temperature anomalies and stratospheric sudden warming
  events.
\newblock {\em Clim. Dyn.} {\bf 2018}, {\em 51},~581--595.

\bibitem[Xie et~al.(2016)Xie, Li, Tian, Fu, Jin, Hu, Zhang, Wang, Sun, Feng,
  et~al.]{xie2016connection}
Xie, F.; Li, J.; Tian, W.; Fu, Q.; Jin, F.F.; Hu, Y.; Zhang, J.; Wang, W.; Sun,
  C.; Feng, J.;  et~al.
\newblock A connection from Arctic stratospheric ozone to El Ni{\~n}o-Southern
  oscillation.
\newblock {\em Environ. Res. Lett.} {\bf 2016}, {\em 11},~124026.

\bibitem[Gray et~al.(2010)Gray, Beer, Geller, Haigh, Lockwood, Matthes,
  Cubasch, Fleitmann, Harrison, Hood, et~al.]{gray2010solar}
Gray, L.J.; Beer, J.; Geller, M.; Haigh, J.D.; Lockwood, M.; Matthes, K.;
  Cubasch, U.; Fleitmann, D.; Harrison, G.; Hood, L.;  et~al.
\newblock Solar influences on climate.
\newblock {\em Rev.  Geophys.} {\bf 2010}, {\em \hl{48}
}.  https://doi.org/10.1029/2009RG000282.

\bibitem[Komitov(2024)]{komitov2024possible}
Komitov, B.
\newblock About the Possible Solar Nature of the $\sim$200 yr (de Vries/Suess)
  and $\sim$2000--2500 yr (Hallstadt) Cycles and Their Influences on the
  Earth’s Climate: The Role of Solar-Triggered Tectonic Processes in General
  “Sun--Climate” Relationship.
\newblock {\em Atmosphere} {\bf 2024}, {\em 15},~612.

\bibitem[Komitov and Kaftan(2022)]{komitov2022danjon}
Komitov, B.; Kaftan, V.
\newblock Danjon Effect, Solar Activity, and Volcanism.
\newblock {\em Geomagn. Aeron.} {\bf 2022}, {\em 62},~1117--1122.

\bibitem[Komitov and Kaftan(2023)]{komitov2023lower}
Komitov, B.; Kaftan, V.
\newblock The Lower Ionosphere and Tectonic Processes on Earth.
\newblock {\em Geomagn. Aeron.} {\bf 2023}, {\em 63},~1038--1046.

\bibitem[Khain and Khalilov(2009)]{khain2009possible}
Khain, V.; Khalilov, E.
\newblock About possible influence of solar activity on seismic and volcanic
  activities: Long-term forecast.
\newblock {\em Sci. Without Borders} {\bf 2009}, {\em \hl{316}
}.

\bibitem[QU et~al.(2011)QU, Huang, Du, ZHAO, DENG, and Cao]{qu2011periodicity}
Qu, W.Z.; Huang, F.; Du, L.; Zhao, J.P.; Deng, S.G.; Cao, Y.
\newblock The periodicity of volcano activity and its reflection in some
  climate factors.
\newblock {\em Chin. J. Geophys.} {\bf 2011}, {\em 54},~135--149.

\bibitem[Yamazaki and Nakamura(2021)]{yamazaki2021stratospheric}
Yamazaki, K.; Nakamura, T.
\newblock The stratospheric QBO affects antarctic sea ice through the tropical
  convection in early austral winter.
\newblock {\em Polar Sci.} {\bf 2021}, {\em 28},~100674.

\bibitem[Rao et~al.(2023)Rao, Garfinkel, Ren, Wu, Lu, and
  Chu]{rao2023projected}
Rao, J.; Garfinkel, C.I.; Ren, R.; Wu, T.; Lu, Y.; Chu, M.
\newblock Projected strengthening impact of the Quasi-Biennial Oscillation on
  the Southern Hemisphere by CMIP5/6 models.
\newblock {\em J. Clim.} {\bf 2023}, {\em 36},~5461--5476.

\bibitem[Butler and Domeisen(2021)]{butler2021wave}
Butler, A.H.; Domeisen, D.I.
\newblock The wave geometry of final stratospheric warming events.
\newblock {\em Weather. Clim. Dyn.} {\bf 2021}, {\em 2},~453--474.

\bibitem[Zambri et~al.(2021)Zambri, Solomon, Thompson, and
  Fu]{zambri2021emergence}
Zambri, B.; Solomon, S.; Thompson, D.W.; Fu, Q.
\newblock Emergence of Southern Hemisphere stratospheric circulation changes in
  response to ozone recovery.
\newblock {\em Nat. Geosci.} {\bf 2021}, {\em 14},~638--644.

\bibitem[Zuev and Savelieva(2019)]{zuev2019cause}
Zuev, V.V.; Savelieva, E.
\newblock The cause of the spring strengthening of the Antarctic polar vortex.
\newblock {\em Dyn. Atmos. Oceans} {\bf 2019}, {\em
  87},~101097.

\end{thebibliography}
\end{document}